\documentclass[fleqn,usenatbib]{mnras}

\usepackage{newtxtext,newtxmath}

\usepackage[T1]{fontenc}

\DeclareRobustCommand{\VAN}[3]{#2}
\let\VANthebibliography\thebibliography
\def\thebibliography{\DeclareRobustCommand{\VAN}[3]{##3}\VANthebibliography}
\newcommand{\dotdeg}{\rlap{.}^\circ}
\setcitestyle{notesep={ }} 

\usepackage{graphicx}	
\usepackage{amsmath}	
\usepackage{lipsum}
\usepackage{multirow}
\usepackage[round-mode=places,
            separate-uncertainty=true,
            table-align-uncertainty=true]{siunitx_3_3_17}
\usepackage[table]{xcolor}

\title[Radio dipole: NVSS \& RACS]{A Bayesian approach to the cosmic dipole in radio galaxy surveys: \\
Joint analysis of NVSS \& RACS}

\author[Oayda et al.]{
Oliver T. Oayda,$^{1}$\thanks{E-mail: oliver.oayda@sydney.edu.au}
Vasudev Mittal$^{1, 2}$,
Geraint F. Lewis$^{1}$, 
and Tara Murphy$^{1}$
\\
$^{1}$Sydney Institute for Astronomy, School of Physics A28, The University of Sydney, NSW 2006, Australia\\
$^{2}$Department of Physical Sciences, IISER Mohali, Knowledge City, Sector 81, SAS Nagar, Manauli PO 140306, Punjab, India\\
}

\date{Accepted XXX. Received YYY; in original form ZZZ}

\pubyear{2024}

\begin{document}
\label{firstpage}
\pagerange{\pageref{firstpage}--\pageref{lastpage}}
\maketitle

\begin{abstract}
We examine the sky distribution of radio galaxies in the NRAO VLA Sky Survey (NVSS) and the Rapid ASKAP Continuum Survey (RACS).
Analyses of these samples have reported tension between their inferred dipoles and the kinematic dipole of the Cosmic Microwave Background (CMB).
This represents a challenge to the traditional assumption that the Universe is homogeneous and isotropic on large scales: the cosmological principle.
We find that NVSS and RACS contain local radio sources which give a non-negligible contribution to the overall dipole signal.
These need to be adequately accounted for since the aim is to probe the composition of the Universe at large scales.  
By appropriately considering these sources,
the inferred dipole amplitude in either sample is reduced.
Nonetheless, we find support for a dipole aligning with that of the CMB
but larger in amplitude, especially in the joint analysis.
However, the `clustering dipole' --
the contribution of local sources to the net inferred dipole --
appears to align with the direction of the CMB dipole,
and its magnitude increases as deeper nearby sources are considered up to a comoving distance of $\approx 130$ Mpc ($h=0.7$).
The significance of this observation in the context of the cosmological principle is unclear and prompts further inquiry.
\end{abstract}




\section{Introduction}
A critical assumption in modern cosmology -- the Lambda cold dark matter ($\Lambda$CDM) paradigm -- is that the Universe is homogeneous and isotropic on large scales.
Such a postulate about the symmetry of space was initially made by Einstein,
but it was later termed the `cosmological principle' (CP) by Milne \citep{milne1935, peebles2020}.
The CP is implicitly incorporated into the Friedmann-Lema\^{i}tre-Robertson-Walker (FLRW) spacetime metric,
greatly simplifying its form.
Now, the CP can be supported by the smoothness of the Cosmic Microwave Background (CMB),
although one must additionally couple this empirical observation
with the metaphysical assumption that all observers in the Universe perceive the same CMB.

Small-scale CMB temperature fluctuations, generated from matter-photon interactions before they decoupled, are of order $\Delta T / T \approx 10^{-5}$.
Imprinted on this smooth temperature map is a dipole of order $\Delta T / T \approx 10^{-3}$ (the `kinematic dipole'),
which is normally interpreted as arising from the Earth's peculiar motion towards $(l,b) = (264\dotdeg021, 48\dotdeg253)$ in Galactic coordinates at a speed of $v_{\text{CMB}} = 369.82\pm0.11\,\text{km}\,\text{s}^{-1}$ \citep{planck2020}.
But since the small-scale anisotropies are thought to be the progenitors of structures in the late universe under $\Lambda$CDM,
the dipole-subtracted frame must be one in which the cosmological principle holds.
In this frame, distributions of matter in the Universe should not have a preferred direction of higher or lower density.
However, where the Earth's peculiar motion has not been factored out, distributions of distant matter should exhibit a dipole (the `matter dipole') aligning in direction and magnitude with the CMB temperature dipole.
If this is not the case, this suggests against the applicability of the CP to observational data, and hence to our prevailing understanding of the Universe more generally.

In recent studies, especially those using large catalogues of radio galaxies covering significant fractions of the sky, the matter dipole and CMB dipole appear to align in direction but not magnitude, with the matter dipole being in excess \citep[see e.g. the reviews of][]{Peebles_2022, aluri2023}.
In this work, we turn to two radio catalogues that have previously provided evidence for this magnitude discrepancy, namely the Rapid Australian SKA Pathfinder Continuum Survey \citep[RACS;][]{racs-original} at 887.5 MHz \citep[RACS-low;][]{racs-low} and the National Radio Astronomy Observatory Very Large Array Sky Survey \citep[NVSS;][]{nvss-survey}.
We examine both catalogues with a joint Bayesian methodology, probing the anisotropic distribution of their radio galaxies and comparing it to the CMB dipole.

This paper is structured as follows.
In Section~\ref{sec:background}, we give an overview of the state of the literature relating to tests of the CP and the matter dipole studies.
In Section~\ref{sec:nvss-racs}, we present the raw NVSS and RACS-low samples and our approach to preparing them for analysis, and in Section~\ref{sec:approach}, we explain our method of statistical analysis.
We give our results in Section~\ref{sec:results} and discuss them in Section~\ref{sec:discussion-conclusion}, in which we conclude with our salient finding.

\section{Background}
\label{sec:background}
The CP forms the backbone of our cosmological framework, hence its empirical verification is extremely important.
If the assumption of isotropy and homogeneity is correct, then the thermal dipole of the CMB is explained by the Earth's peculiar motion, which imprints a Doppler modulation on an otherwise isotropic CMB.
An independent way of verifying this claim is to search for a dipolar modulation in surveys of distant sources such as radio galaxies and quasars.
The Earth's motion is anticipated to induce this dipole, and it should be consistent with the CMB dipole in direction and magnitude if the cosmological principle holds.
In this vein, \citet{ellis1984} proposed a number count dipole test of the CP.
They assumed the following:
\begin{enumerate}
    \item An observer is moving with a velocity much less than $c$ with respect to a homogeneous and isotropic distribution of distant sources.
    \item The apparent flux density $S$ can be described by a cumulative power law, where $N(>S) \propto S^{-x}$.
    \item Within the observer's passband, the spectral energy distribution for sources has a power law dependence on frequency $S_f \propto f ^{-\alpha}$ for spectral index $\alpha$ and frequency $f$.
\end{enumerate}
Under these assumptions, relativistic aberration and Doppler boosting will modulate the distribution of sources in the observer's frame, inducing a dipole anisotropy with amplitude 
\begin{equation}
        {\bf \mathcal{D}} = [2 + x(1+\alpha)]\frac{v}{c}. \label{eq:dipole-magnitude}
\end{equation}
Thus, the modulation in number density observed for a patch of sky containing $N$ sources and pointing in some direction $\bf \hat{n}$ will be
\begin{equation}
        \frac{\Delta N}{N} = {\bf D}\cdot{\bf \hat{n}} = [2 + x(1+\alpha)]\frac{\bf v}{c}\cdot{\bf \hat{n}}. \label{eq:E&B-dipole}
\end{equation}
\citet{ellis1984} supposed that $O(10^{5})$ sources would be needed to observe this modulation.
It is worth mentioning that some of the assumptions made by the original authors have been questioned.
For instance, \citet{dalang2022} \citep[see also][]{gaundalin2023}
have claimed that the flux distribution exponent $x$
as well as the spectral index $\alpha$ are redshift-dependent,
which impacts the resulting dipole measure
\citep[although][has suggested the dipole measure
is robust to these effects]{vonHausegger2024}.
One other outstanding issue is the assumption that the inferred dipole is solely contributed to by the Earth's peculiar motion, which is a central theme in this work.
In general, the inhomogeneous distribution of sources in the Earth's local neighbourhood can contribute to the overall dipole signal.
This is an undesired effect, since the aim of the \citet{ellis1984} test is to probe the distant background of sources where the CP is expected to hold -- that is, where the Universe is expected to average out to homogeneity and isotropy.
For example, \citet{tiwari2016} suggest that the bulk of sources analysed need to be at a high redshift ($z \approx 1$) to avoid local source contamination.
We will return to this point below.

The \citet{ellis1984} test has to date been widely performed with a number of different samples.
One significant first result was that of \citet{blake2002}, in which the authors examined NVSS and reported broad agreement (within uncertainties) of the NVSS radio galaxy dipole and the kinematic dipole.
However, this stands as an isolated result; many subsequent studies found a discrepancy between the matter dipole and the CMB dipole.
We outline a handful of the key results here.
\citet{singal2011} reported that the NVSS dipole is consistent in direction with the CMB dipole, but about 4 times larger than expected in magnitude.
The trend of an unexpectedly large NVSS amplitude continued in subsequent works \citep[see e.g.][]{gibelyou2012, rubart2013, colin2017, bengaly2018, siewert2021, singal2023b, singal2023, wagenveld2023}.
A number of these studies used additional radio catalogues beyond NVSS (such as RACS for the later works), usually arriving at similar conclusions.
The weight of these studies has lead to the prevailing sentiment that the matter dipole is consistent with the CMB dipole in direction but anomalously larger in magnitude \citep{snowmass2022, aluri2023}.

More recently, \citet{secrest2021} expanded the scope of the matter dipole studies from radio sources to quasars.
Analysing the CatWISE2020 catalogue \citep{marocco2021}, the authors reported the presence of an anomalously large matter dipole at a $4.9 \sigma$ statistical significance.
This finding was followed up in \citet{secrest2022} with a joint analysis of CatWISE2020 and NVSS, confirming an excessive dipole amplitude at a $5.1 \sigma$ joint statistical significance.
However,
\citet{abghari2024} has recently questioned the CatWISE2020 result,
pointing to peculiarities regarding the ecliptic bias in the sample,
as well as coupling between the dipole mode and higher order multipoles
-- which impacts the dipole estimator used in \citet{secrest2021}.

We note that, in general, the radio galaxy and quasar studies rely on frequentist statistical techniques, using estimators for calculating the dipole.
Estimators are sometimes characterised by an inherent bias.
As an example, \citet{siewert2021} illustrate that the linear estimator used in earlier studies has a directional bias depending on which part of the sky is masked.
In simulations, this has been shown to draw the inferred dipole away from the direction of the true dipole.
In contrast to frequentist methods, the tools of Bayesian statistics have not been used as extensively. 
Bayesian techniques were implemented in \citet{dam2023}'s study of CatWISE2020, which supported the findings of \citet{secrest2021}.
However, \citet{mittal2024} found that an independent catalogue -- the Quaia quasar sample \citep{storeyfisher2023quaia} -- was consistent with the kinematic dipole, although it was noted that the sample suffers from contamination near the Galactic plane which needs to be accounted for.

Apart from the matter dipole studies, many other methods to check for isotropy have been proposed. 
\citet{aluri2023} gives an extensive discussion on these tests, but some of them involve probing Type Ia SNe \citep[see e.g.][]{horstmann2022, singal2022, sorrenti2022} and the FLRW metric by testing spatial curvature \citep[see e.g.][]{zhou2020}.
Recently \citet{oayda2023} proposed a method of calculating the dipole in the time dilation of sources with intrinsic time-scales.

The final point to consider is the effect of local clustering, which we alluded to earlier.
Some studies make the assumption that, on account of most sources being at high redshifts, its effects will be negligible \citep[see e.g.][for a recent contribution analysing NVSS and RACS-low]{wagenveld2023}.
Other studies have anticipated the degree to which clustering will affect the overall dipole signal, using theoretical arguments from $\Lambda$CDM to construct a prior likelihood for a `clustering dipole' term \citep[see e.g.][]{tiwari2016, cheng2023, dam2023}.
Since the matter dipole amplitude is consistently in excess across many independent studies, it is worth probing the fraction of the amplitude that can be explained by local clustering, if at all.
This is especially the case given the significance that a genuine tension would have for the CP and modern cosmology.
Thus, we revisit two of the catalogues which have seemingly contributed to the tension: NVSS and RACS-low.
Although many studies have calculated the matter dipole across a spectrum of different samples,
their analysis is generally limited to characterising the dipole amplitude and direction
and evaluating its significance with respect
to a null hypothesis of no dipole signature.
In this work, we jointly analyse NVSS and RACS-low, utilising Bayesian
inference to compare the relative strengths of competing hypotheses
by their marginal likelihoods and
to understand if the inferred matter dipole is in agreement with the CMB kinematic dipole.

\section{Radio Galaxy Samples}
\label{sec:nvss-racs}
At the first stage of processing, both the NVSS and RACS-low source catalogues are binned into equal-area sky pixels using the \textsc{healpix}\footnote{\url{https://healpix.sourceforge.io/}} procedure implemented in the \textsc{python} package \textsc{healpy} \citep{Gorski2005, Zonca2019}.
We set the $N_\text{side}$ parameter to 64, generating $49\,152$ pixels with an angular size of about 55 arcminutes.

\subsection{NVSS}
\subsubsection{Flux limit and mask}
The NVSS survey was conducted using the Very Large Array (VLA) at 1.4 GHz, between 1993 to 1997 \citep{nvss-survey}.
It covers about 82\% of the sky, above declination $\delta \geq -40^\circ$).
The full source catalogue consists of 1.8 million sources.
In this work, we make use of the integrated flux densities recorded for each source.

\begin{figure}
    \includegraphics[width=\columnwidth]{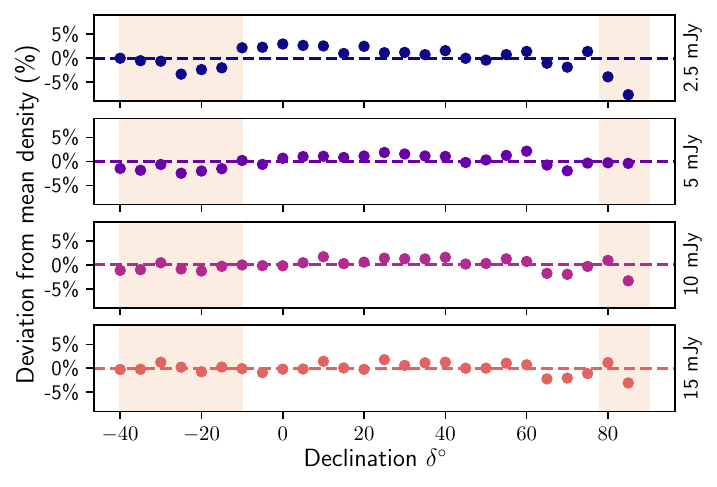}
    \caption{Deviation from NVSS(B) mean source density (\%) by declination angle for different flux density cuts. The declinations using the DnC configuration are shown in the shaded light brown region, and the D configuration region is left unshaded.
    The flux density cuts chosen are labelled on the right of each panel.}
    \label{fig:nvss_density}
\end{figure}

\begin{figure}
    \includegraphics[width=\columnwidth]{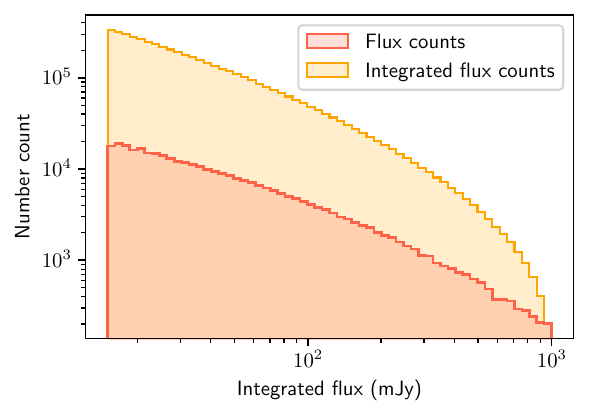}
    \vspace{-2.5mm}
    \caption{Flux density distribution of the NVSS(B) sample, shown in red, in which a 15 mJy flux limit has been used. Overlaid onto this in yellow is the integrated flux density distribution i.e. the total number of sources above a limiting flux density.}
    \label{fig:nvss_flux}
\end{figure}

NVSS used two configurations of the VLA for different sky regions.
Namely, the D configuration covered declinations $-10^\circ \leq \delta \leq 78^\circ$, with the DnC configuration covering the remaining portion of the sky.
Critically, this means that without a sufficiently bright flux density cut,
NVSS shows a strong systematic bias in number density with declination -- even though NVSS is claimed to reach 100\% completeness by about 4 mJy \citep{nvss-survey}.
While some authors have chosen to mitigate this with a 10 mJy flux density cut \citep[see e.g.][]{secrest2022}, we instead select a 15 mJy limit in line with for example \citet{tiwari2016} and \citet{wagenveld2023}.
We illustrate the effect of declination on source density in Fig.~\ref{fig:nvss_density}.
Moving from lower to higher flux density cuts (top to bottom), the average source density by declination becomes increasingly clustered around the mean density.
From 2.5 mJy to 15 mJy, the average percentage deviation drops from $\approx 1.8\%$ to $\approx 0.9 \%$.
We also show the resulting flux density distribution
with the 15 mJy cut applied in Fig.~\ref{fig:nvss_flux}.
Note that both Fig.~\ref{fig:nvss_density} and Fig.~\ref{fig:nvss_flux} are generated after incorporating masking and source selection, described below.

NVSS suffers from significant contamination near the Galactic plane, appearing as localised regions of high source density.
We therefore choose to mask Galactic latitudes $|b| \leq 10^\circ$.
However, we also identified other regions of spuriously high source counts.
Accordingly, we follow some of the masking choices identified in \citet{cheng2023}.
First, bright, extended sources can appear as multiple entries in the NVSS catalogue, so we
impose an upper flux density limit of 1~Jy,
such that the flux density distribution of
our NVSS sample spans
$15\,\text{mJy} \leq S \leq 1000\,\text{mJy}$.
As an additional measure,
one might choose to mask all pixels containing a bright source
with flux greater than 1~Jy, or alternatively construct a circular
mask centred around that bright source \citep[see e.g.][]{blake2002a}.
We tested this, but ultimately found it had not effect on our results.
Therefore, we persist with the upper flux density cutoff.
We then visually identify a number of localised regions with substantially higher number counts and mask these regions with discs of
varying radii.
We show the Galactic coordinates of these regions in the corresponding rows of Table~\ref{tab:disc_masks}.
A number of known bright radio sources are located in or at the edge of these masked discs, which we specify in the `A-source' column of the same table.
Where there is no relevant nearby source, the cell is left blank.
\begin{table}
    \centering
    \begin{tabular}{l S[round-precision=1, table-format=3.1] S[round-precision=1, table-format=2.1] S[round-precision=1, table-format=1.1] c}
         \hline Sample & {$l^\circ$} & {$b^\circ$} & {$r^\circ$} & A-source \\\hline
        NVSS    & 239.9 & -56.6 & 1.1 & Fornax A   \\
                & 207.8 & -18.6 &2.9 & Orion A    \\
                & 286.2 & 73.2  & "   & Virgo A    \\
                & 21.7  & 19.8  & 1.1 & --    \\
                & 291.1 & 64.6  & "   & --    \\
                & 119.1 & -19.2 & "   & --    \\
                & 156.0 & 20.2  & "   & --    \\\hline
        RACS-low& 107.0 & -43.0 & 3.4 & --         \\
                & 79.0  & -30.0 & "   & --         \\
                & 25.0  & 26.0  & "   & Hercules A \\
                & 286.7 & 73.5  & "   & Virgo A    \\
                & 280.8 & -33.1 & "   & LMC    \\
                & 184.8 & -5.5  & "   & Crab    \\
                \hline
    \end{tabular}
    \caption{Centre position in galactic coordinates $(l^\circ, b^\circ)$ and radius $r^\circ$ of the distinct disc masks used in NVSS and RACS-low. Bright radio sources contained within or at the edge of the discs are indicated in the `A-source' column.}
    \label{tab:disc_masks}
\end{table}
Lastly, we mask an additional degree north of the NVSS declination cutoff to remove pixels with lower source counts at the survey limit.
Our continuous region includes $\delta > 40^\circ$, leaving a final count of 341,072 NVSS sources.
The main features of the mask can be seen in the left panes of Fig.~\ref{fig:sky_samples}.

\subsubsection{Local clustering}
\label{sub:nvss_clustering}
It is conceivable that the inhomogeneous distribution of low redshift sources will contribute to the recovered dipole signal in NVSS.
Since the cosmological principle is a statement about the composition of the Universe on the largest scales, ideally any contamination from nearby clusters should be avoided.
As mentioned, some studies \citep[see e.g.][dealing with CatWISE2020 and NVSS respectively]{dam2023, cheng2023} account for this possibility by explicitly determining prior likelihood functions for the clustering signal, taking into account predictions from $\Lambda$CDM.
In this case, the net inferred dipole would include a contribution from a clustering term.

Here, we investigate the effect of clustering by cross-matching NVSS with known local radio sources, creating two variant catalogues: one with local sources included, and one in which local sources have been removed.
The first is constructed as we have already described above.
To construct the second, similar to \citet{cheng2023} we use the source catalogue from the Two Micron All Sky Survey Redshift Survey \citep[2MRS;][]{2mrs} and radio sources with redshift less than 0.01 identified on the NASA/IPAC Extragalactic Database (NED).\footnote{\url{https://ned.ipac.caltech.edu/}}
The positional accuracy of NVSS sources with flux density greater than 15 mJy is $\lesssim 1$ arcsecond.
We therefore adopt a cross-matching radius of 5 arcseconds to capture as many genuine local sources as possible while minimising spurious matches.
Specifically, for each local source, we find all NVSS sources within 5 arcseconds and remove the one with the smallest positional offset.
We apply the cross-matching regime before masking but after flux-limiting the sample.
With the 15 mJy cut, 3049 sources are removed from the NVSS catalogue.
After masking, this final sample has 338,222 NVSS sources, 2850 less than the sample with local sources included.
For future reference, we refer to sample with local sources removed as NVSS(B), and the sample containing local sources as NVSS(A).

\subsection{RACS-low}
\subsubsection{Flux limit and mask}
The RACS-low survey \citep{racs-low} was conducted with ASKAP at a central frequency of 887.5~MHz.
Observations took place between 2019 and 2020.
The survey covered declinations $\delta \leq 41^\circ$.
The resolution of RACS-low changes with declination, and so in addition to the main catalogue, a catalogue constructed from data convolved to a common resolution of 25 arcseconds was produced.
This is the catalogue we used in this work.
However, not all observational tiles could be convolved to the common resolution, and so the effective coverage of the catalogue used here is a continuous region between $-80^\circ \leq \delta \leq 30^\circ$, excluding the Galactic plane for $|b| < 5^\circ$.
This covers about 67.9\% of the sky, and the catalogue contains about 2.1 million sources.

\begin{figure}
    \includegraphics[width=\columnwidth]{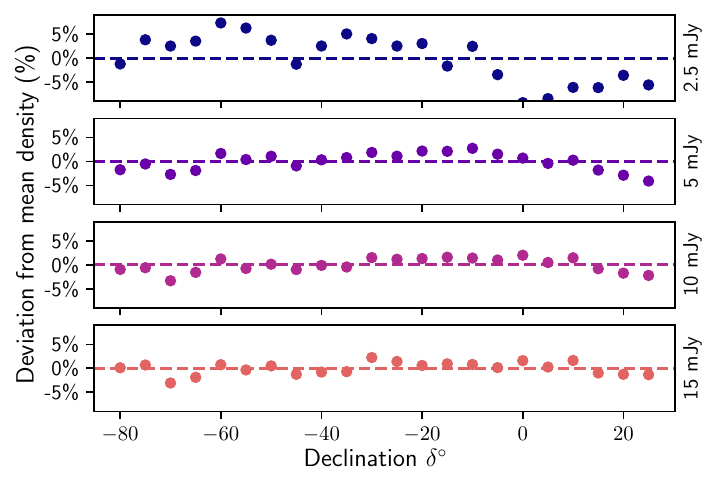}
    \caption{Deviation from RACS-low(B) mean source density (\%) by declination angle for different flux density cuts.
    The flux density cuts chosen are labelled on the right of each panel.}
    \label{fig:racs_density}
\end{figure}

\begin{figure}
    \includegraphics[width=\columnwidth]{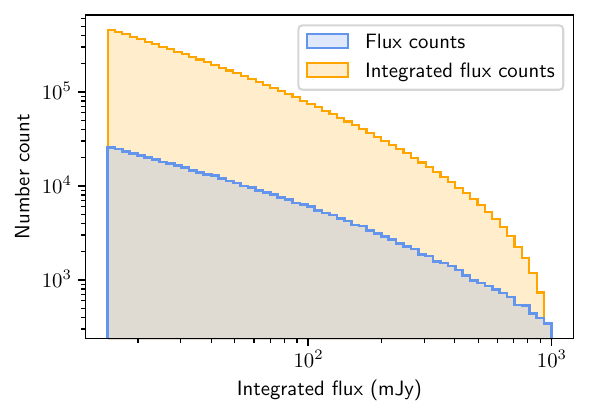}
    \vspace{-3.5mm}
    \caption{Flux density distribution of the RACS-low(B) sample, shown in blue, in which a 15 mJy flux limit has been used. Overlaid onto this in yellow is the integrated density flux distribution i.e. the total number of sources above a limiting flux density.}
    \label{fig:racs_flux}
\end{figure}

We first probe the impact of our choice of limiting flux density.
All our references to flux with respect to RACS-low refer to the integrated flux density, as was the case for NVSS.
RACS-low is asserted to be 95\% complete at about 3 mJy, though we note improvements in the homogeneity of the sample with deeper flux cuts up to about 15 mJy.
To see this, we plot how the source density changes with declination for each flux limit in Fig~\ref{fig:racs_density}.
From 2.5 mJy to 15 mJy, the average percentage deviation from the mean density decreases from about 4.3\% to 1.1\%.
We ultimately use a 15 mJy cut for our final sample,
and, as for NVSS,
use an upper flux density limit of 1000 mJy.
We also show the sample flux distribution in Fig.~\ref{fig:racs_flux}.
Note that both Fig.~\ref{fig:racs_density} and Fig.~\ref{fig:racs_flux} are made after the masking and source selection that we describe below.

Turning to the choice of mask, we remove under-dense pixels along the edge of the excluded Galactic plane.
We also mask the excluded survey tiles at the southern pole with a disc of radius $13^\circ$, and mask an additional degree of declination in the northern hemisphere such that our final sample covers declinations $-77^\circ < \delta < 29^\circ$.
Finally, we note localised regions with a dearth of source counts which persist in our sample, corresponding to under-dense tiles in the original RACS-low merged sky catalogue.
These can be seen in Figure 4 of \citet{racs-low} near the upper declination limit of the survey.
Similar to our approach for NVSS, we mask these regions with discs of radius $\approx 3^\circ$.
The position, exact radius and any known nearby bright radio sources are specified in the corresponding rows of Table~\ref{tab:disc_masks}.
We also mask an over-dense region proximate
to the LMC, as shown in the same table.
This leaves us with a sample of
462,911 RACS-low sources.
The key features of the mask can be seen in the right panes of Fig.~\ref{fig:sky_samples}.

\subsubsection{Local clustering}
We repeat the same cross-matching procedure as for NVSS (see Section~\ref{sub:nvss_clustering}).
The typical RACS-low source position is accurate to within about 1 to 2 arcseconds.
For this reason, we keep the same cross-matching radius of 5 arcseconds as with NVSS.
This identifies 3700
local objects in RACS-low between the 2MRS and NED catalogues, which we remove, leaving a final source count after masking of 459,276.
We use the same naming convention as above, referring to the sample with local sources removed as RACS-low(B), and the variant containing local sources as RACS-low(A).

\subsection{Summary}
\begin{figure*}
    \includegraphics[width=\columnwidth]{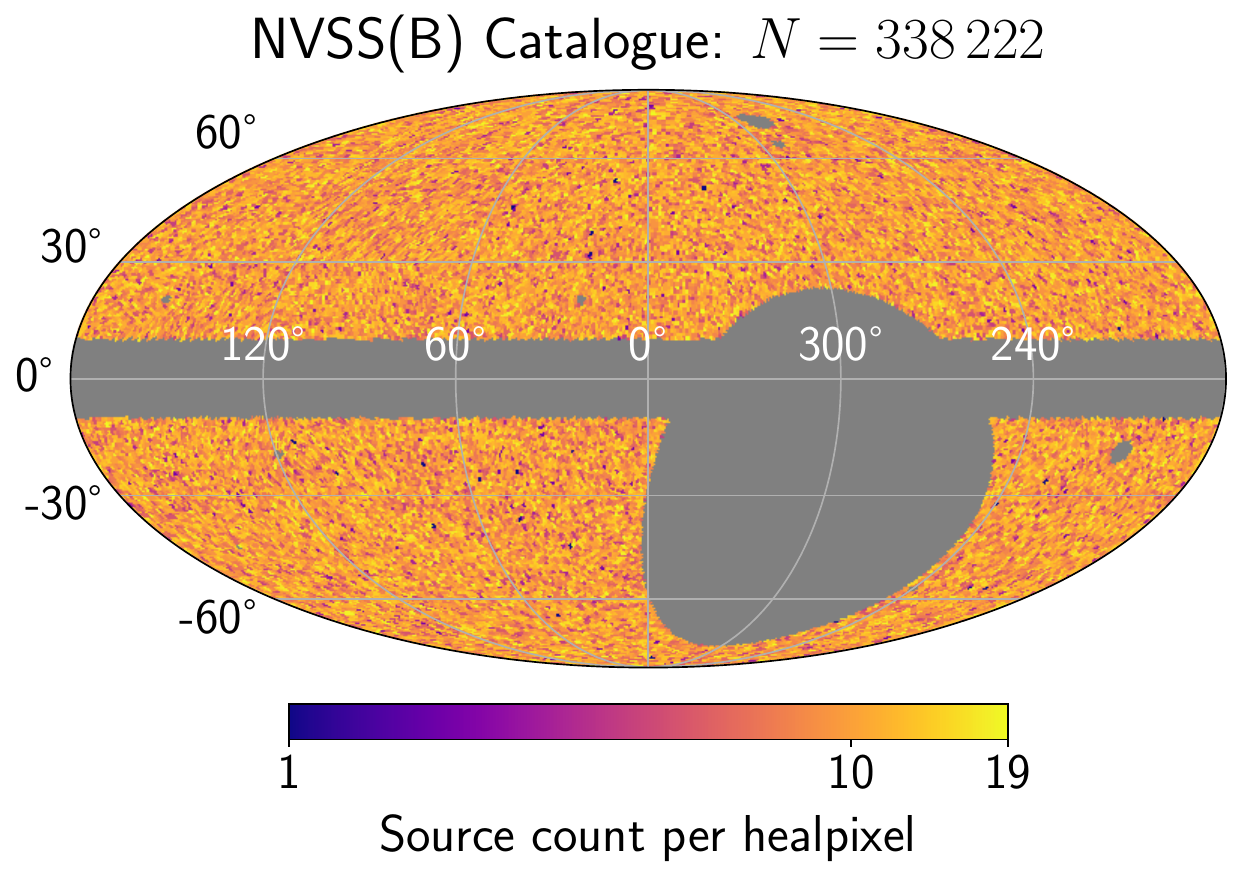}
    \hfill
    \includegraphics[width=\columnwidth]{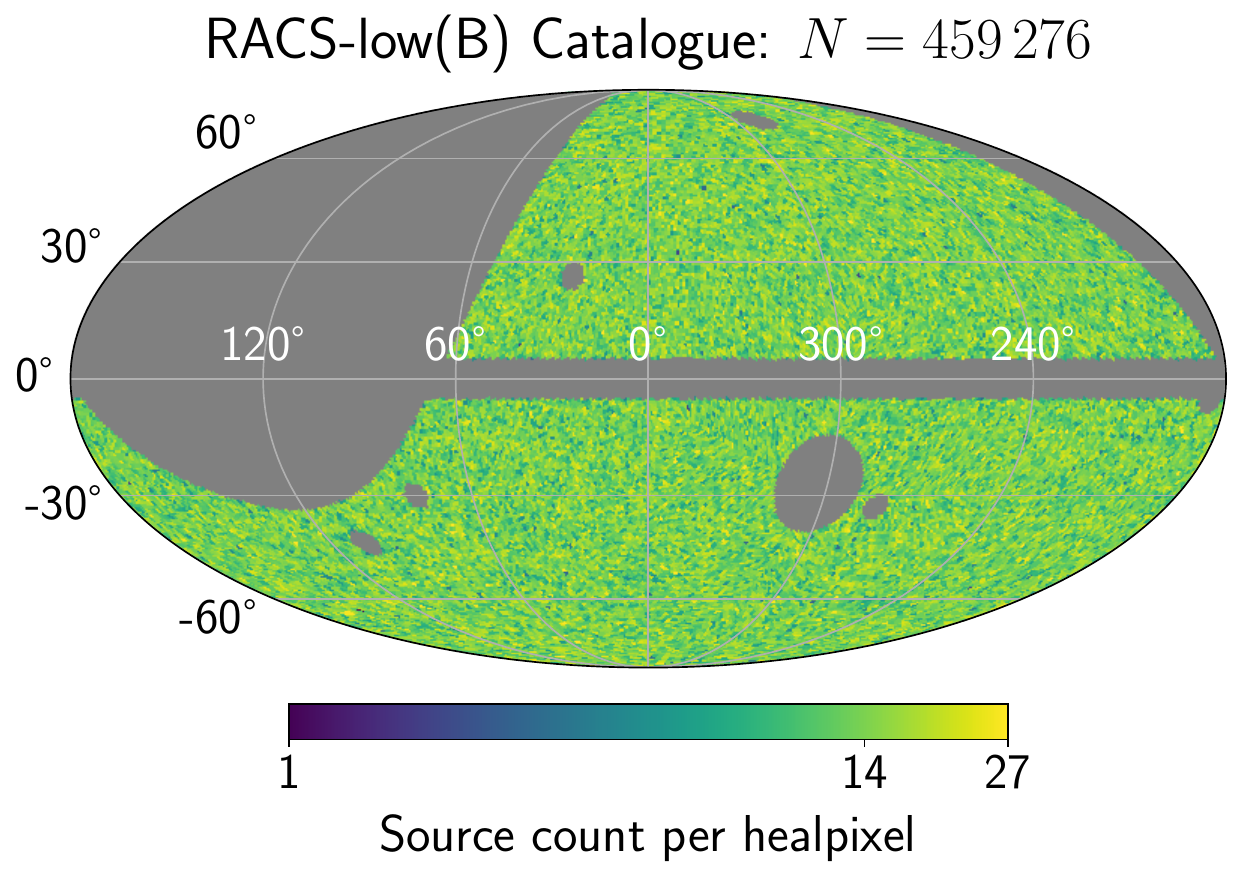}
    \includegraphics[width=\columnwidth]{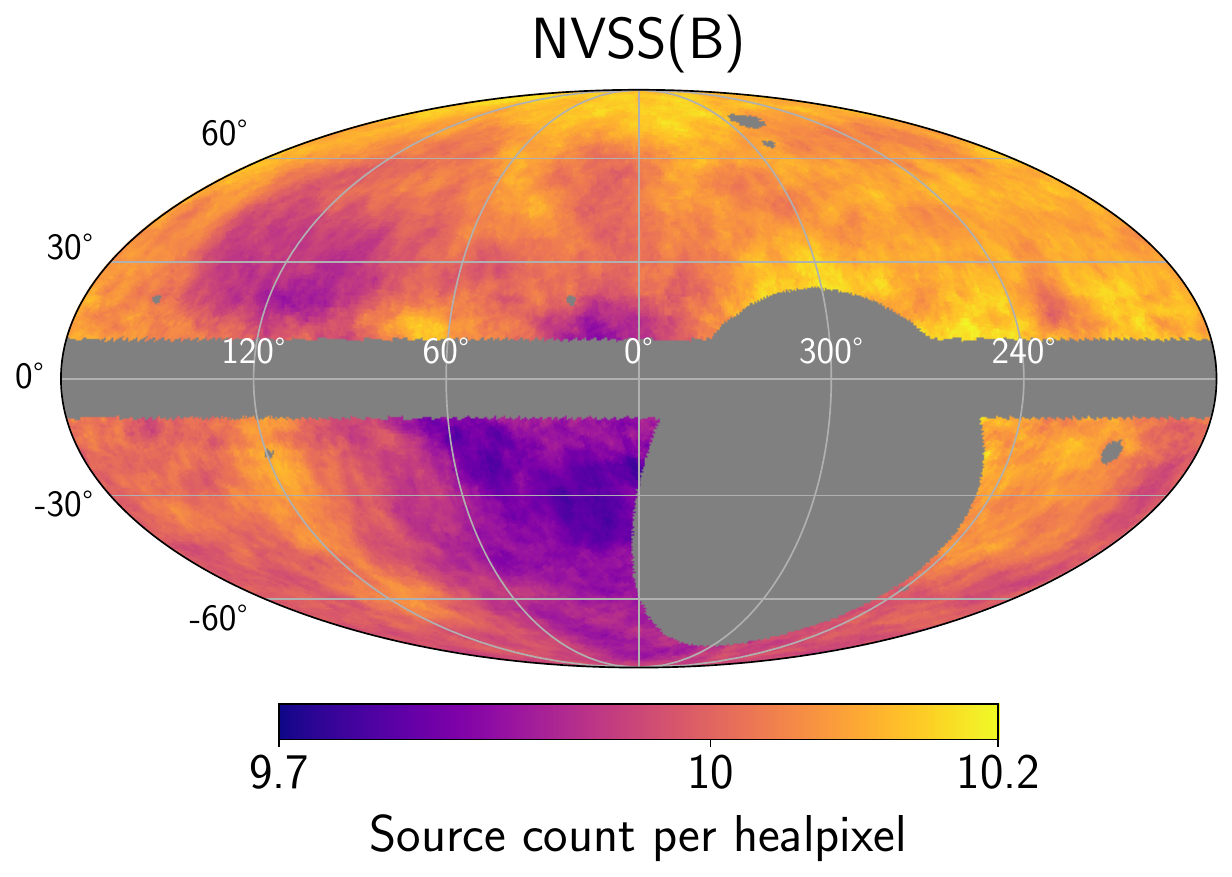}
    \hfill
    \includegraphics[width=\columnwidth]{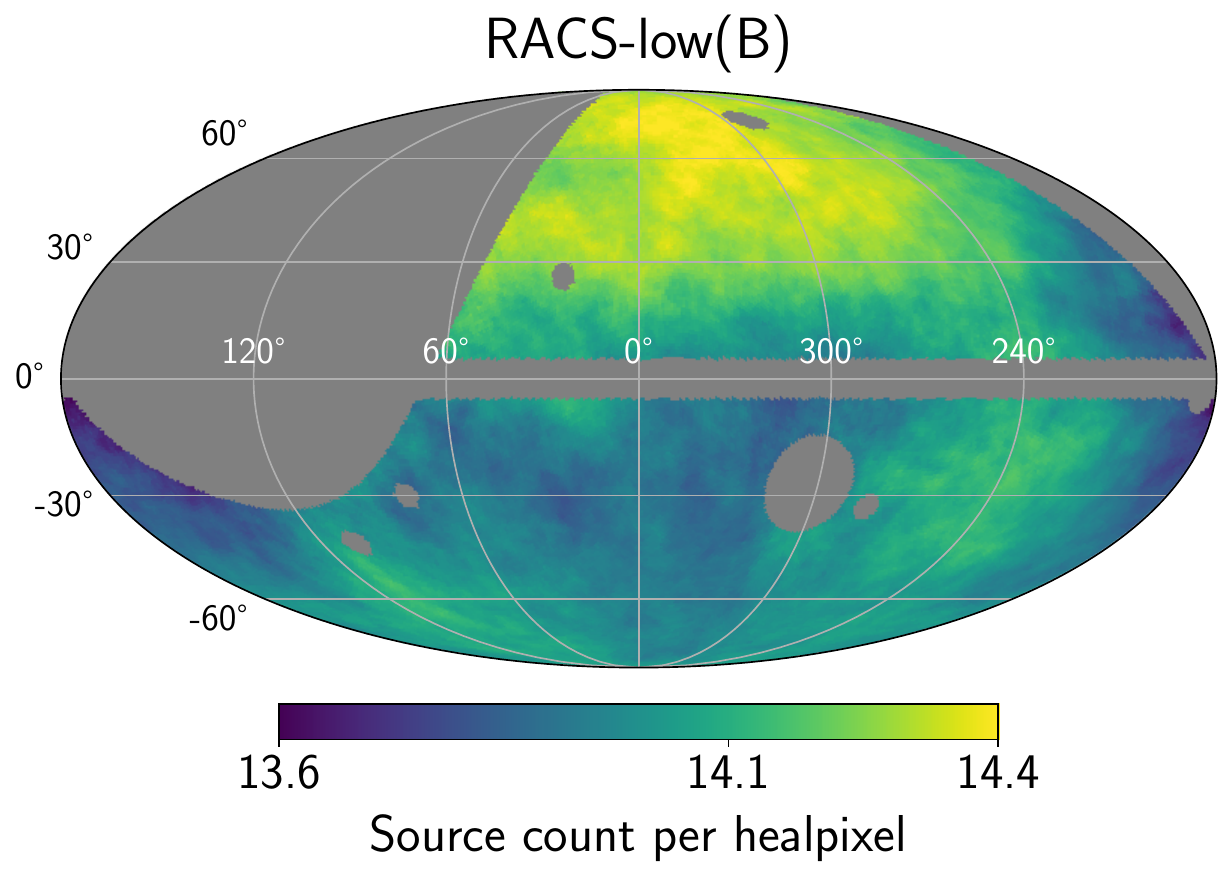}
    \caption{Sky projection (Mollweide) of our final cross-matched (B) samples
    in Galactic coordinates,
    binned into healpixels ($N_\text{side} = 64$; pixel angular size of 55').
    Masked regions are indicated in grey. \textit{Left:} NVSS(B). \textit{Right:} RACS-low(B). \textit{Top row:} Number counts for each pixel. The mean source count is given by the middle tick of the colour bar. \textit{Bottom row:} Averaged counts over a 1 steradian scale. The mean over all averaged counts is given by the middle tick of the colour bar.}
    \label{fig:sky_samples}
\end{figure*}

We give the sky projections of our final samples in the top row of Fig.~\ref{fig:sky_samples}.
Note that there we have plotted the B variants of NVSS and RACS-low (the cross-matched samples), since with a visual inspection there are only very minor differences between the A and B variants.
We also show a smoothed version of the raw samples in the bottom row of the same figure.
Here, for each unmasked pixel, we compute the average density across all pixels within 1 steradian.
We also give a summary of the key choices we made in processing both catalogues in Table~\ref{tab:samples} to aid the reader.
\begin{table}
    \centering
    \setlength{\tabcolsep}{3pt}
    \begin{tabular}{lcccc}
        \hline
         Sample     & Flux range (mJy)      & $\delta$ included    & Gal. mask          & $N$      \\
         \hline
         NVSS(A)    & $15 \leq S \leq 1000$ & $\delta > -40^\circ$ & $|b| \leq 10^\circ$& 341\,072 \\
         NVSS(B)    & "                     & "                    & "                  & 338\,222 \\\hline
         RACS-low(A)& $15 \leq S \leq 1000$ & $-77^\circ < \delta < 29^\circ$ & $|b| \leq 5^\circ$ & 462\,911 \\
         RACS-low(B)& "                      & "                              & "                  & 459\,276 \\\hline
    \end{tabular}
    \caption{Summary of final RACS-low and NVSS samples, including flux cuts, masking choices and total number of sources. (A) samples are the catalogues containing local sources, and (B) samples have had local sources removed.}
    \label{tab:samples}
\end{table}

\section{Approach}
\label{sec:approach}

\subsection{Expected amplitude}
\label{sub:expected_amp}
The conventional approach to finding the expected matter dipole amplitude is to use the spectral indices of the sources $\alpha$ and the slope of the flux density distribution $x$ at the low flux density limit.
These terms are then used in the expression of \citet{ellis1984}
-- see equation~\eqref{eq:dipole-magnitude} --
with $v \approx 369$~km\,s$^{-1}$, as determined from the CMB temperature dipole.

\subsubsection{Spectral index}
Since both the RACS-low and NVSS catalogues list flux densities at a single frequency, we do not at first instance have information about each source's spectral index.
Typically, the expected spectral index for synchrotron emission -- the assumed emission mechanism -- is used \citep[$\alpha \approx 0.75$; see e.g.][]{secrest2022}.
Often, overlapping sources between radio surveys at different frequencies can be used to infer the spectral index distribution, the median of which is the origin of the typical value of $\alpha \approx 0.75$.
For example, Figure 8 of \citet{summs} gives the distribution of spectral indices for NVSS sources overlapping with the Sydney University Molonglo Sky Survey (SUMSS), showing that the distribution is reasonably broad about the median of $\approx 0.83$.
Although we find that the expected amplitude is not very sensitive to the choice of $\alpha$, to safeguard against the possibility, we can assume that the median spectral index is 0.75 but has a significant spread.
We take the largest dispersion from Figure 8, being about 0.5, and assume that the distribution of NVSS spectral indices can be described by a Gaussian $\mathcal{G}(\mu=0.75, \sigma=0.5)$.
For RACS-low, \citet{racs-low} found $\alpha$ distributions through comparison with four other surveys at different frequencies, including NVSS.
The typical $\alpha$ was between 0.6 and 0.9, with dispersions at most being $\approx 0.5$, and we therefore use the same Gaussian distribution to describe the RACS-low spectral indices.

\subsubsection{Source number counts}
\begin{figure}
    \includegraphics[width=\columnwidth]{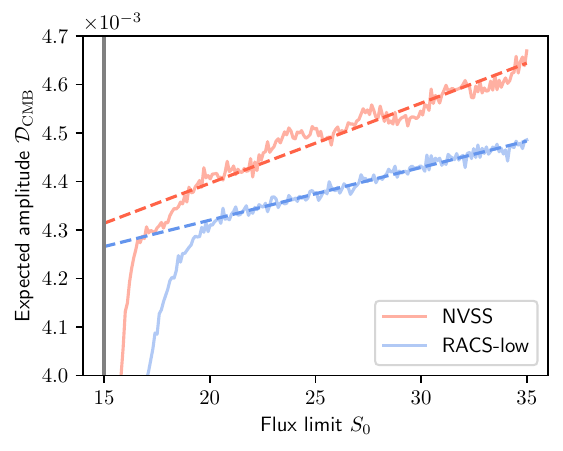}
    \caption{Computed expected amplitude $\mathcal{D}_{\text{CMB}}$
    for NVSS and RACS-low with different choices of $S_0$.
    The desired value of $\mathcal{D}_{\text{CMB}}$ is the extrapolated
    intercept at $S_0 = 15$ mJy, as indicated by the dashed line.}
    \label{fig:expected_amp}
\end{figure}

While we could use the slope of the flux density distribution $x$
to ultimately arrive at the expected amplitude,
we instead follow the approach developed in \citet{mittal2024} but with modifications to allow flux density measurement uncertainties to be adequately accounted for.
Namely, we randomly regenerate either catalogue a number of times using the recorded flux density and its associated uncertainty, as we describe below.
\begin{enumerate}
    \item For each original NVSS or RACS-low source $i$ with flux density $S_i$,
        we generate a new flux value $S_i^*$ by drawing from a Gaussian
        $\mathcal{G}(\mu = S_i, \sigma = \Delta S_i)$
        for flux density measurement uncertainty $\Delta S_i$.
    \item For each source $i$,
        we draw a spectral index $\alpha_i$ from the Gaussian
        $\mathcal{G}(\mu=0.75, \sigma=0.5)$.
    \item We then determine the number of new sources $n_i$
        above some limiting flux density $S_0$.
    \item For each source $i$, we apply a Doppler shift $S^*_i \delta^{1+\alpha_i}$
        for $\delta = \gamma (1 + \beta \cos \theta)$
        with $\beta = v_{\text{CMB}}/c$ and $\theta = 0$, as in \citet{ellis1984}.
    \item We then find the number of boosted sources $n_b$
        above the same $S_0$ and multiply that number by $\delta^2$,
        accounting for relativistic aberration.
    \item The expected amplitude is then determined via
        \begin{equation}
            \mathcal{D} = \frac{n_b - n_i}{n_i}.
        \end{equation}
\end{enumerate}
For a given $S_0$,
we can repeat (i)--(vi) a number of times to arrive at a
mean and standard deviation for the expected dipole amplitude.
However, if we choose $S_0$ to be too close to the actual flux density
limit we impose for NVSS and RACS-low (15 mJy),
then the dipole amplitude is underestimated.
This can be seen by the drop-off near
$S_0 = 15$ mJy in Fig.~\ref{fig:expected_amp}.
The drop-off in dipole amplitude is
a boundary effect caused by the manner in which we resample flux densities.
To see this, suppose we pick $S_0 = 25$ mJy.
A source flux density slightly below this limit
has a non-negligible likelihood of being resampled above the limit,
owing to its measurement uncertainty.
This is averaged over the many runs of our procedure.
But, if we choose $S_0 \approx 15$ mJy,
sources below 15 mJy cannot be resampled above the limit since,
by construction,
they are excluded from either catalogue.
Meanwhile,
the dipole amplitude is expected to rise with
$S_0$, as seen in Fig.~\ref{fig:expected_amp},
since the NVSS and RACS-low cumulative flux density distributions become steeper
for larger limiting flux densities.

We can circumvent this issue by extrapolating the gradient
in $\mathcal{D}_{\text{CMB}}$ versus $S_0$ back to 15 mJy,
as seen by the dashed line in Fig.~\ref{fig:expected_amp}.
We do so by running our procedure (steps i--vi) 100 times
for different values of $S_0$ between 15 mJy and 35 mJy,
computing the mean dipole amplitude over those runs.
We then determine the gradient
using a straightforward least-squares linear fit
for $S_0 > 17$ mJy (NVSS) and $S_0 > 20$ mJy (RACS-low).
Extrapolating back to $S_0 = 15$ mJy, we arrive at
$\mathcal{D}_{\text{CMB}} = 4.31 \times 10^{-3}$ for NVSS and
$\mathcal{D}_{\text{CMB}} = 4.27 \times 10^{-3}$ for RACS-low.
Note that in the procedure described above,
the B/cross-matched sample variant is used;
the change in amplitude using the A variant is negligible.

\subsection{Bayesian inference}
\subsubsection{Overview}
Our statistical approach is similar to and explained in depth in \cite{mittal2024}, but we recapitulate the main concepts here.
The process of Bayesian inference can be discretised into two stages: the level of parameter optimisation and the level of model inference \citep{mackay2003}.
In the first stage, we wish to populate the terms of Bayes's theorem and solve for the posterior distribution of a model $M$'s parameters, where
\begin{equation}
    P( \mathbf{\Theta} | \mathbf{D}, M ) = \frac{\mathcal{L}(\mathbf{D} | \mathbf{\Theta}, M) \pi(\mathbf{\Theta} | M)}{\mathcal{Z}(\mathbf{D} | M)} \label{eq:bayes-theorem}
\end{equation}
for data $\mathbf{D}$ and model parameters $\mathbf{\Theta}$, as well as likelihood, prior and marginal likelihood functions $\mathcal{L}$, $\pi$ and $\mathcal{Z}$ respectively \citep[as using the notation of][]{speagle2020}.

At the second stage, models are ranked according to the marginal likelihood, an integral over all parameter space: $\mathcal{Z} = \int_{\Omega_{\mathbf{\Theta}}} \mathcal{L}(\mathbf{\Theta}) \times \pi(\mathbf{\Theta}) \, d \mathbf{\Theta}$.
This works by reframing Bayes' theorem as the posterior probability for
a model directly, such that
\begin{equation}
    P(M | \mathbf{D}) = \frac{\mathcal{L}(\mathbf{D} | M) \pi(M)}{\mathcal{Z}(\mathbf{D})}.
\end{equation}
Thus, noting that the term $\mathcal{Z}(\mathbf{D})$ cancels,
the ratio of posterior probabilities for two models is
\begin{equation}
    \frac{P(M_i | \mathbf{D})}{P(M_j | \mathbf{D})}
        = \frac{\pi(M_i)}{\pi(M_j)}
        \frac{\mathcal{L}(\mathbf{D} | M_i)}{\mathcal{L}(\mathbf{D} | M_j)}
        = \frac{\pi(M_i)}{\pi(M_j)}
        B_{ij}
\end{equation}
where $\mathcal{L}(\mathbf{D} | M)$ is identical to the marginal likelihood $\mathcal{Z}$ of equation~\eqref{eq:bayes-theorem} and $B_{ij}$ is the Bayes factor: the ratio of model marginal likelihoods.
$\pi(M)$ is identified as the model prior likelihood, representing our beliefs about the relative strengths of each model before any knowledge of the data (see Section~\ref{subs:priors}).
Accordingly,
in this work we compare models using the natural logarithm of the Bayes factor,
using the null hypothesis as a common benchmark.
For example, if models $M_1$ and $M_2$ have marginal likelihood $\mathcal{Z}_1$ and $\mathcal{Z}_2$, and the null hypothesis has marginal likelihood $\mathcal{Z}_0$, we can effectively rank the explanatory power of $M_1$ and $M_2$ by comparing $\ln B_{10} = \mathcal{Z}_1 - \mathcal{Z}_0$ and $\ln B_{20} = \mathcal{Z}_2 - \mathcal{Z}_0$.
This is advantageous where there are many models and it would be cumbersome to directly write down the Bayes factors between all possible pairs of models.
The degree to which one Bayes factor is larger than the other is a reflection of the extent to which a model better explains the data over others while conserving prior volume.
A qualitative interpretation (the `Jeffrey's scale')
of the Bayes factors can be found in \citet{kass1995};
we use those normative categories in this work,
although we caution that they use units of $2 \ln B$,
whereas here all Bayes factors are presented as $\ln B$.

What remains is to fill the terms of Bayes's theorem at equation~\eqref{eq:bayes-theorem}.
For each model, we construct a likelihood function and determine the prior likelihoods for its parameters, which we explain below.
Then, parameter optimisation, as well as the calculation of the marginal likelihood, is handled by the \textsc{python} package \textsc{dynesty} \citep{dynesty-v2.1.3}.\footnote{\url{https://pypi.org/project/dynesty/}}
\textsc{dynesty} implements the Nested Sampling algorithm to efficiently sample the posterior distribution in shells of increasing likelihood, also allowing the marginal likelihood to be evaluated \citep{skilling2004, skilling2006}.

\subsubsection{Likelihood functions}
In \citet{mittal2024}, two types of likelihood functions were tested alongside each other: the Poissonian approach and the point-by-point approach.
Both were found to give consistent results, and so in this work we chose to use only the point-by-point case, mainly since it is computationally less expensive.
Under this approach, we consider some arbitrary function $f = f(\mathbf{\hat{p}}_i)$ which gives a scalar for the $i$-th pixel on the sky as described by the unit vector $\mathbf{\hat{p}}_i$ pointing to that pixel.
Such a function needs to be normalised where for example arbitrary sky masks are chosen.
Thus, the normalised value of the function at pixel $i$ is
\begin{equation}
    \hat{f}(\mathbf{\hat{p}}_i) = \frac{f(\mathbf{\hat{p}}_i)}{\sum_{i=1}^{n_{\text{pix}}} f(\mathbf{\hat{p}}_i)}
\end{equation}
for total number of pixels $n_{\text{pix}}$.
This is our sky probability map, the form of which depends on the assumptions about the chosen signal: a dipole, monopole, etc.
The likelihood function can then be written as
\begin{equation}
    \ln \mathcal{L} = \sum_{i=1}^{_{\text{pix}}} N_i \ln \hat{f}(\mathbf{\hat{p}}_i) \label{eq:pbp-likelihood}
\end{equation}
for number of sources $N_i$ in pixel $i$.
This works because each source in a pixel is associated with the same function value $\hat{f}$, and so the $i$-th pixel contributes the value $[\hat{f}(\mathbf{\hat{p}}_i)]^{N_i}$ to the likelihood function.
With this established, we need only to consider what the functional form of $f$ will be for each model.
\begin{enumerate}
    \item Starting with the null hypothesis, we suppose that either sample exhibits uniform density over the sky.
    In this case, we need to weight all pixels equally, favouring no part of the sky over other regions.
    This can be done by setting $f$ to 1 for all pixels, and so this model has zero parameters i.e. $\mathbf{\Theta}_{\text{null}} = \emptyset$.
    \item Next, we introduce a dipole vector $\mathbf{D}$ of magnitude $\mathcal{D}$ and pointing in some direction $l$, $b$ in Galactic coordinates.
    The scalar value for an arbitrary pixel on the sky is the sum of the monopole signal and the dipole signal $\mathbf{D} \cdot \mathbf{\hat{p}}_i$.
    That is, we fit a monopole and dipole simultaneously.
    This leads to
    \begin{equation}
        f(\mathbf{\hat{p}}_i) = 1 + \mathcal{D} \cos \theta_i
    \end{equation}
    where $\theta_i$ is the angle between the $i$-th pixel unit vector and the dipole vector.
    Of course, the value of $\theta_i$ depends on the direction of the dipole, which leaves the model parameters as $\mathbf{\Theta}_{\text{dipole}} = \{ \mathcal{D}, l, b \}$.
    \item We may also fix any of the parameters in the dipole model to constant values in order to test alternative hypotheses.
    Specifically, we fix $(l,b) = (264\dotdeg021, 48\dotdeg253)$ -- the direction of the CMB dipole -- and term this the `kinematic direction' model.
    \item We also fix the magnitude $\mathcal{D}$ to either of the expectations determined in Section~\ref{sub:expected_amp} while leaving the direction free, terming this the `kinematic velocity' model.
    \item We additionally investigate the `kinematic dipole' model in which all parameters $D$, $l$ and $b$ are fixed to their CMB-derived values.
    \item Finally, in our joint analysis (described below),
        for one of our models we set the prior likelihood function
        for the model parameters
        to the posterior distribution of \citet{wagenveld2023}.
        We approximate this distribution,
        as based on the information provided in the study,
        by drawing the parameters from Gaussians
        with mean and standard deviation given by the quoted median value and error.
        Namely, we have
        $\mathcal{D}_{\text{NVSS}} = \mathcal{D}_{\text{RACS}}
                \sim \mathcal{G}(\mu= 0.0129,\sigma=0.0018)$,
        $l \sim \mathcal{G}(\mu=269\dotdeg5, \sigma=8^\circ)$ and
        $b \sim \mathcal{G}(\mu=56\dotdeg2, \sigma=11^\circ)$.
        This is the `W23' model.
\end{enumerate}
We summarise the six models mentioned above in Table~\ref{tab:model_reference}, in which we have labelled each model $M_0$ through $M_5$.

\begin{table}
    \centering
    \begin{tabular}{l l l}
         \hline & Short label & Description \\\hline
         $M_0$ & Null  & Monopole i.e. uniform number density \\
         $M_1$ & Free dipole & Dipole with free parameters $\mathcal{D}$, $l$, $b$ \\
         $M_2$ & Kinematic velocity & Dipole with $\mathcal{D}$ fixed and free $l$, $b$ \\
         $M_3$ & Kinematic direction & Dipole with $l$, $b$ fixed and free $\mathcal{D}$ \\
         $M_4$ & Kinematic dipole & All parameters fixed to CMB expectation \\
         $M_5$ & W23 & $\mathcal{D}$, $l$, $b$ from \citet{wagenveld2023} \\\hline
    \end{tabular}
    \caption{Description of the six models used in this work labelled $M_0$ through $M_5$.}
    \label{tab:model_reference}
\end{table}

In this work, we analyse NVSS and RACS-low individually as well as jointly, so we consider now how our approach differs in the joint case.
Some studies, for example \citet{darling2022}, combine multiple radio galaxy catalogues by scaling fluxes according to the source spectral index.
There is the possibility that doing so leads to a spurious dipole signal by neglecting varying systematic effects across both catalogues, as was noted in \citet{wagenveld2023}.
Here, we need not combine RACS-low and NVSS; rather, we suppose each catalogue is influenced by the same dipole signal, although, since they will not have the exact same flux or spectral index distribution, the dipole magnitudes will be different.
That is to say, this joint model has parameters $\mathbf{\Theta} = \{\mathcal{D}_{\text{NVSS}}, \mathcal{D}_{\text{RACS}}, l, b\}$.
We fit a dipole to either catalogue separately, but combine information across both catalogues with the joint likelihood function
\begin{equation}
    \ln \mathcal{L} = \ln \mathcal{L}_{\text{NVSS}} + \ln \mathcal{L}_{\text{RACS}} \label{eq:joint_likelihood}
\end{equation}
where either individual likelihood is determined through equation~\eqref{eq:pbp-likelihood}.

\subsubsection{Prior likelihood functions}
\label{subs:priors}
Our choice of priors for the model parameters $\mathcal{D}$, $l$, $b$ is conditioned on the information we have before any knowledge of the data.
We describe the choices for each in turn.
\begin{enumerate}
    \item For all dipole amplitudes, we sample from a uniform distribution $\mathcal{U}[0, 0.1]$. This reflects the spectrum of different results across independent studies \citep[see e.g.][]{snowmass2022, aluri2023}.
    \item We sample $l$ and $b$ uniformly over the sky, preferring no direction over another. To do this, since internally our calculations are performed in equatorial coordinates, we sample the right ascension in radians according to $\phi \sim \mathcal{U}[0, 2 \pi]$. We then sample the co-declination according to $\theta \sim \cos^{-1} (1 - 2u)$ where $u \sim \mathcal{U}[0,1]$. It is important not to sample the co-declination straightforwardly from a uniform distribution between $0$ and $\pi$ since the area element of a sphere is dependent on the co-latitude; doing so would lead to a bias in the prior likelihood function towards the poles.
    \item For the prior likelihood of each model $\pi(M)$,
              we set each model to be equiprobable; that is, $\pi(M) = 1/5$
              where each sample is analysed individually
              and $\pi(M) = 1/6$ in the joint analysis.
\end{enumerate}

\section{Results}
\label{sec:results}
We present the results for each of our models,
including Bayes factors and posterior distributions where noted.
As a quick reference for the reader,
we also summarise all inferred dipole amplitudes in Table~\ref{tab:amp_table}.

\subsection{Individual catalogues}
\subsubsection{NVSS}
\begin{figure*}
    \centering
    \includegraphics[width=\columnwidth]{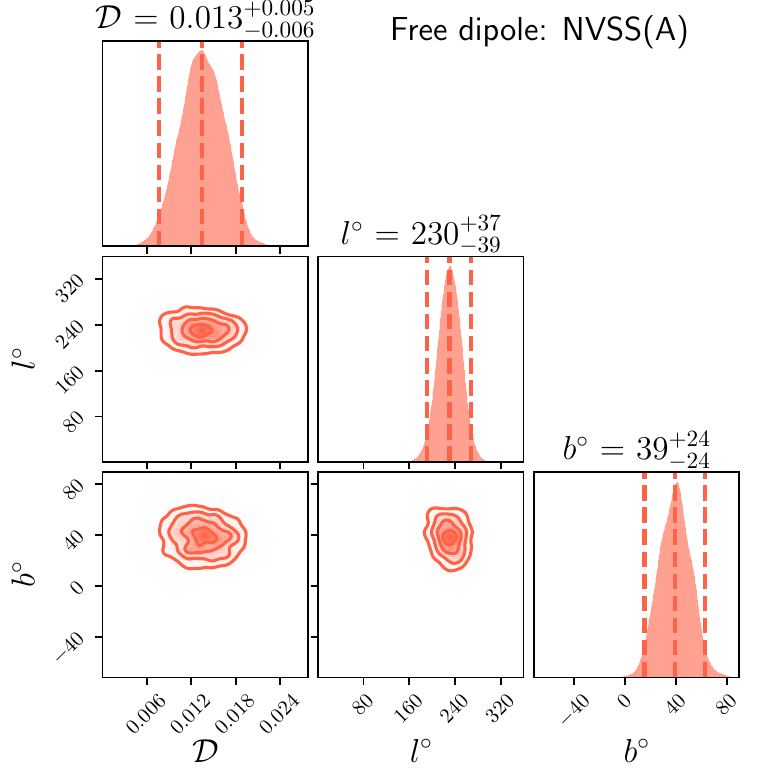}
    \hfill
    \includegraphics[width=\columnwidth]{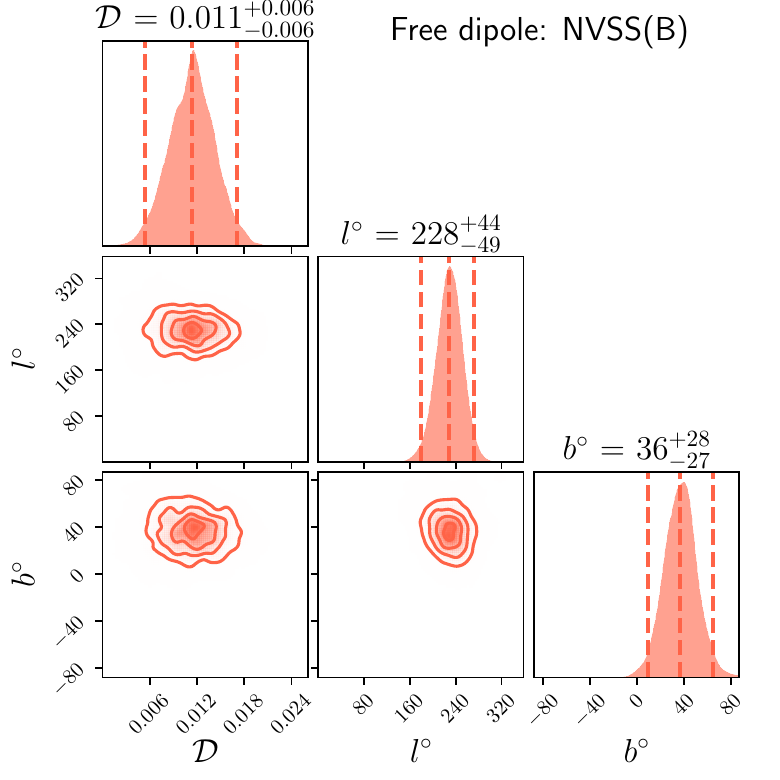}
    \includegraphics[width=\columnwidth]{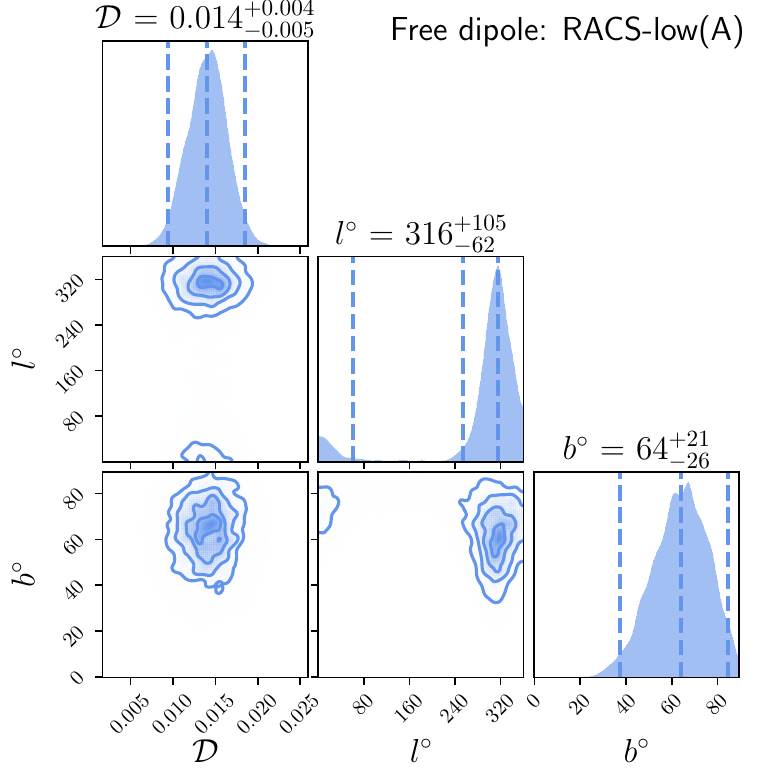}
    \hfill
    \includegraphics[width=\columnwidth]{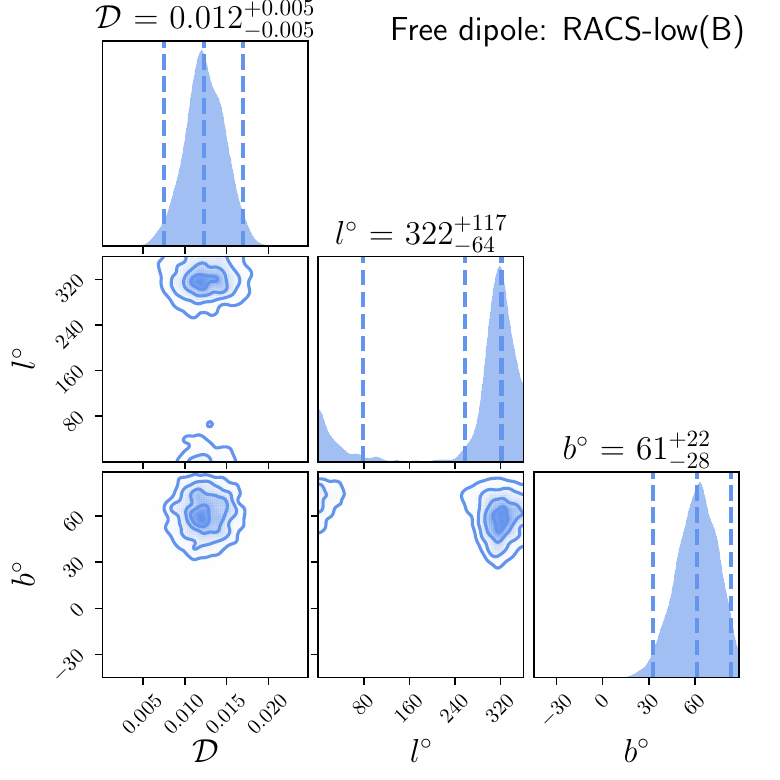}
    \caption{Corner plots for the free dipole model tested on our individual samples.
    The dashed lines on the 1D marginal posteriors represent a 95\% credible interval
    i.e. $2 \sigma$ either side of the median.
    Quoted uncertainties above each 1D posterior also give
    $2 \sigma$ credible limits on each model parameter.
    The contours on the 2D marginal posteriors
    give intervals of $0.5\sigma$,
    encompassing 11.8\%, 39.4\%, 67.5\% and 86.4\% of the distribution.
    \textit{Top row (red)}: NVSS(A) and NVSS(B) samples.
    \textit{Bottom row (blue)}: RACS-low(A) and RACS-low(B) samples.}
    \label{fig:all_corners}
\end{figure*}

We give our Bayes factors by model for both NVSS(A) and NVSS(B) in Table~\ref{tab:nvss_bayes}.
\begin{table}
    \centering
    \begin{tabular}{l l S[round-precision=1]}
        \hline Sample              & Model               & {Log Bayes factor}\\\hline
        NVSS(A)\multirow{5}{*}     & $M_0$ \ \ Null                & {--}             \\
                                   & $M_1$ \ \ Free dipole         & 5.8        \\
                                   & $M_2$ \ \ Kinematic velocity  & 3.7        \\
        \rowcolor{black!10}        & $M_3$ \ \ Kinematic direction & 7.6        \\
                                   & $M_4$ \ \ Kinematic dipole    & 5.9        \\\hline
        NVSS(B)\multirow{5}{*}     & $M_0$ \ \ Null                & {--}             \\
                                   & $M_1$ \ \ Free dipole         & 3.1        \\
                                   & $M_2$ \ \ Kinematic velocity  & 2.8        \\
                                   & $M_3$ \ \ Kinematic direction & 4.5        \\
        \rowcolor{black!10}        & $M_4$ \ \ Kinematic dipole    & 4.8        \\\hline
    \end{tabular}
    \caption{Bayes factors by model for each NVSS sample. The highlighted row indicates which model has the highest Bayes factor -- hence marginal likelihood -- for each sample.}
    \label{tab:nvss_bayes}
\end{table}
For NVSS(A), the model with the greatest explanatory power is the kinematic direction model, which again assumes a dipole aligning with the CMB dipole but free in magnitude.
In this case, we find an amplitude of $\mathcal{D} \approx (12 \pm 5) \times 10^{-3}$ with $2\sigma$ uncertainties.
Although the free dipole is not the most favoured model,
we represent its posterior distribution for the fit to NVSS(A) with a corner plot,
shown in the left pane of Fig.~\ref{fig:all_corners} (top row).
This is largely for illustration,
indicating the dipole signature that is picked up at the first level of inference.
Note that the inferred amplitude of the free dipole ($M_1$)
is not necessarily the same as that of the kinematic direction model ($M_3$),
since these are distinct hypotheses.
We show the free dipole fits for all subsequent samples as well.

For NVSS(B), we instead find that the kinematic dipole model best explains the data, which assumes a dipole totally aligning with the CMB dipole in direction and magnitude.
This is illustrated in the corresponding highlighted row of Table~\ref{tab:nvss_bayes}.
However, the difference in marginal likelihoods between this model and the next best explanation -- the kinematic direction model -- is 0.3 log units.
This is only worth a `bare mention of support' for the kinematic dipole \citep{kass1995}.
Again, we show the corner plot for the free dipole fit on NVSS(B) in the right pane of Fig.~\ref{fig:all_corners} (top row).
This illustrates that the free dipole model infers an amplitude of $\mathcal{D} \approx (11 \pm 6) \times 10^{-3}$.
Notably, the median dipole amplitude has diminished by $\approx 2 \times 10^{-3}$ moving from the NVSS(A) to NVSS(B) free dipole fit, although the uncertainties remain large.
Lastly, if we fix the direction of the NVSS(B) dipole to that of the CMB ($M_3$),
we find that $\mathcal{D} \approx (10 \pm 5) \times 10^{-3}$.

\subsubsection{RACS-low}
We give the Bayes factors for both RACS-low samples in Table~\ref{tab:racs_bayes}.
\begin{table}
    \centering
    \begin{tabular}{l l S[round-precision=1]}
        \hline Sample              & Model               & {Log Bayes factor}\\\hline
        RACS-low(A)\multirow{5}{*} & $M_0$ \ \ Null                & {--}             \\
        \rowcolor{black!10}        & $M_1$ \ \ Free dipole         & 11.9        \\
                                   & $M_2$ \ \ Kinematic velocity  & 6.5        \\
                                   & $M_3$ \ \ Kinematic direction & 11.6       \\
                                   & $M_4$ \ \ Kinematic dipole    & 8.1        \\\hline
        RACS-low(B)\multirow{5}{*} & $M_0$ \ \ Null                & {--}             \\
        \rowcolor{black!10}        & $M_1$ \ \ Free dipole         & 8.2        \\
                                   & $M_2$ \ \ Kinematic velocity  & 5.4        \\
                                   & $M_3$ \ \ Kinematic direction & 7.5        \\
                                   & $M_4$ \ \ Kinematic dipole    & 6.6        \\\hline
    \end{tabular}
    \caption{Bayes factors by model for each RACS-low sample. The highlighted row indicates which model has the highest Bayes factor -- hence marginal likelihood -- for each sample.}
    \label{tab:racs_bayes}
\end{table}
For both RACS-low(A) and RACS-low(B),
the prevailing model is
a free dipole i.e. a dipole where $\mathcal{D}$, $l$, and $b$
are free parameters ($M_1$).
We give the posterior distributions for the free dipole -- shown via corner plots -- for both samples in the bottom row of Fig.~\ref{fig:all_corners}.
Similar to the case of NVSS,
the free dipole amplitude for RACS-low decreases moving from the A to B variant by $\approx 2 \times 10^{-3}$.
For the A variant, we find a free dipole amplitude of
$\mathcal{D} \approx 14\substack{+4 \\ -5} \times 10^{-3}$,
and for the for the B variant,
we find $\mathcal{D} \approx (12 \pm 5) \times 10^{-3}$.

A dipole fixed to the kinematic direction but with free amplitude ($M_3$)
has the second-highest level of support across both variants.
In this case,
we infer $\mathcal{D} \approx 13\substack{+4 \\ -5} \times 10^{-3}$ for the A variant
and $\mathcal{D} \approx 11\substack{+4 \\ -5} \times 10^{-3}$ for the B variant.

\subsection{Joint analysis}
We give the Bayes factors for our joint analysis in Table~\ref{tab:joint_bayes}.
\begin{table}
    \centering
    \begin{tabular}{l l S[round-precision=1]}
        \hline Sample                       & Model               & {Log Bayes factor}   \\\hline
        \multirow{3}{25mm}{NVSS(A) + RACS-low(A)}& $M_0$ \ \ Null           & {--}       \\
                                            & $M_1$ \ \ Free dipole         & 16.1 \\
                                            & $M_2$ \ \ Kinematic velocity  & 11.1 \\
                                            & $M_3$ \ \ Kinematic direction & 19.2 \\
                                            & $M_4$ \ \ Kinematic dipole    & 14.0 \\
        \rowcolor{black!10}                 & $M_5$ \ \ W23                 & 24.5 \\\hline
        \multirow{3}{25mm}{NVSS(B) + RACS-low(B)}& $M_0$ \ \ Null           & {--}       \\
                                            & $M_1$ \ \ Free dipole         & 9.8  \\
                                            & $M_2$ \ \ Kinematic velocity  & 8.8  \\
                                            & $M_3$ \ \ Kinematic direction & 12.1 \\
                                            & $M_4$ \ \ Kinematic dipole    & 11.4 \\
        \rowcolor{black!10}                 & $M_5$ \ \ W23                 & 17.1 \\\hline
    \end{tabular}
    \caption{Bayes factors by model for the joint NVSS(B) + RACS-low(B) sample. The highlighted row indicates which model has the highest Bayes factor -- hence marginal likelihood -- for each sample.}
    \label{tab:joint_bayes}
\end{table}
Here, the prevailing explanation is the
W23 model ($M_5$),
with $\ln B_{50} = 24.5$ where both sample A variants are used
and $\ln B_{50} = 17.1$ where both sample B variants are used.
Compared to the kinematic dipole model,
the W23 model has a log Bayes factor of
$\ln B_{54} = 10.5$ for the A variants and
$\ln B_{54} = 5.7$ for the B variants.
This is beyond overwhelming support for the model
with prior likelihood given by the results of \citet{wagenveld2023}
over the kinematic dipole model,
as using the scale of \citet{kass1995}.
With this prior likelihood,
we infer $\mathcal{D}_{\text{NVSS}} = (12 \pm 3) \times 10^{3}$
and $\mathcal{D}_{\text{RACS}} = (13 \pm 3) \times 10^{3}$
for the sample A variants, as well as
$\mathcal{D}_{\text{NVSS}} = (12 \pm 3) \times 10^{3}$
and $\mathcal{D}_{\text{RACS}} = (12 \pm 3) \times 10^{3}$
for the sample B variants.

The second-most favoured model in either case is the kinematic direction model.
With this model, we find $\mathcal{D}_{\text{NVSS}} = (12 \pm 5) \times 10^{-3}$
and $\mathcal{D}_{\text{RACS}} = 12\substack{+5 \\ -4} \times 10^{-3}$
where the A variants are used,
and $\mathcal{D}_{\text{NVSS}} = (10 \pm 5) \times 10^{-3}$
and $\mathcal{D}_{\text{RACS}} = 11\substack{+5 \\ -4} \times 10^{-3}$
where the B variants are used.

\begin{figure} 
    \includegraphics[width=\columnwidth]{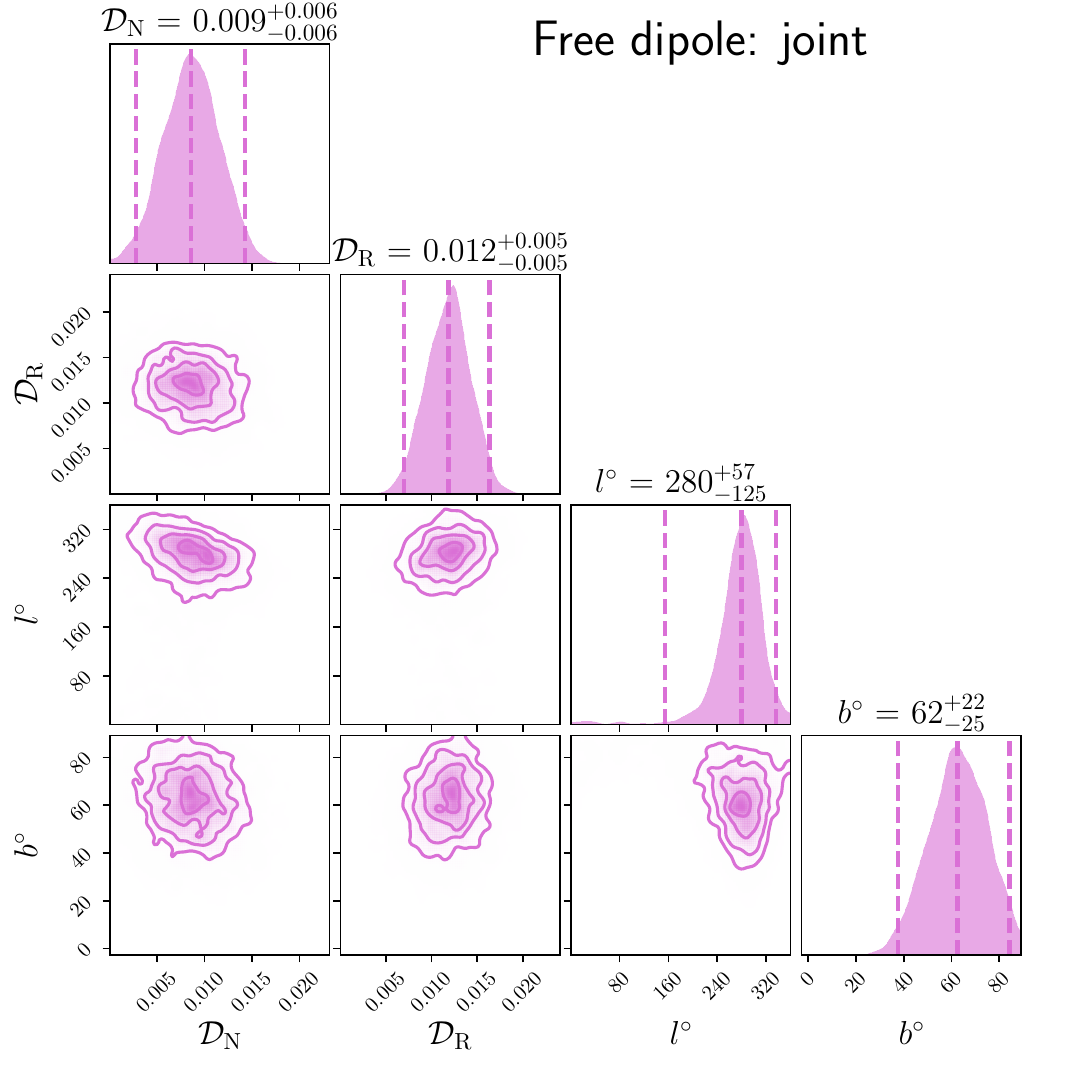}
    \caption{Corner plot for the joint NVSS(B) and RACS-low(B) analysis.
    Although the free dipole is not the strongest model,
    we show its posterior distribution here for illustration.
    The dashed lines represent a 95\% credible interval,
    and the quoted uncertainties above each 1D posterior
    also give $2 \sigma$ credible intervals on each model parameter.
    The contours on the 2D marginal posteriors give intervals of $0.5\sigma$,
    encompassing 11.8\%, 39.4\%, 67.5\% and 86.4\% of the distribution.
    $\mathcal{D}_{\text{N}}$ refers to the NVSS dipole amplitude
    ($\mathcal{D}_{\text{NVSS}}$),
    and $\mathcal{D}_{\text{R}}$ refers to the
    RACS-low dipole amplitude ($\mathcal{D}_{\text{RACS}}$).}
    \label{fig:joint_corner}
\end{figure}

\begin{figure}
    \centering
    \includegraphics[width=\columnwidth]{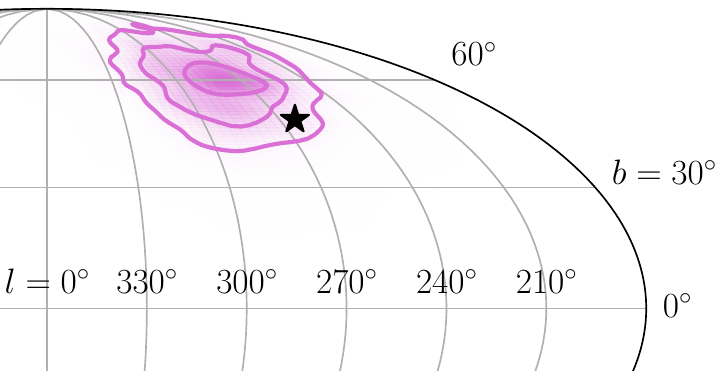}
    \caption{Projection of the marginal posterior distribution ($l$ and $b$) for the joint analysis onto the sky (Mollweide). The contours give intervals of $0.5\sigma$, encompassing 11.8\%, 39.4\% and 67.5\% of the distribution. The black star indicates the direction of the CMB dipole.}
    \label{fig:joint_sky}
\end{figure}

Lastly, we show the model $M_1$ corner plot for the joint analysis with the B variants
in Fig.~\ref{fig:joint_corner},
and project the marginal distributions for $l$ and $b$ onto the sky in Fig.~\ref{fig:joint_sky}.
There, we find $\mathcal{D}_{\text{NVSS}} = (9 \pm 6) \times 10^{-3}$ and $\mathcal{D}_{\text{RACS}} = (12 \pm 5) \times 10^{-3}$.
In the case of the joint analysis with the A variants,
for the free dipole fit we find $\mathcal{D}_{\text{NVSS}} = (11 \pm 6) \times 10^{-3}$ and $\mathcal{D}_{\text{RACS}} = (13 \pm 5) \times 10^{-3}$.

\begin{table}
    \def\arraystretch{1.1}
    \centering
    \begin{tabular}{c l S[round-precision=1] S[round-precision=1]}
        \hline Sample              & Model                    &  {$\mathcal{D}_{\text{NVSS}}$} & {$\mathcal{D}_{\text{RACS}}$} \\\hline
        NVSS(A)\multirow{2}{*}     & $M_3$ \ \ Kin. direction & 12(5)   & {--}    \\
                                   & $M_1$ \ \ Free dipole    & 13(5:6) & {--}    \\
        NVSS(B)\multirow{2}{*}     & $M_3$ \ \ Kin. direction & 10(5)   & {--}    \\
                                   & $M_1$ \ \ Free dipole    & 11(6)   & {--}    \\\hline
        RACS-low(A)\multirow{2}{*} & $M_3$ \ \ Kin. direction & {--}    & 13(4:5) \\
                                   & $M_1$ \ \ Free dipole    & {--}    & 14(4:5) \\
        RACS-low(B)\multirow{2}{*} & $M_3$ \ \ Kin. direction & {--}    & 11(4:5) \\
                                   & $M_1$ \ \ Free dipole    & {--}    & 12(5)   \\\hline
        \multirow{2}{25mm}{
  \centering NVSS(A) + RACS-low(A)}& $M_3$ \ \ Kin. direction & 12(5)   & 12(5:4) \\
                                   & $M_1$ \ \ Free dipole    & 11(6)   & 13(5)   \\
                                   & $M_5$ \ \ W23            & 12(3)   & 13(3)   \\
        \multirow{2}{25mm}{
  \centering NVSS(B) + RACS-low(B)}& $M_3$ \ \ Kin. direction & 10(5)   & 11(5:4) \\
                                   & $M_1$ \ \ Free dipole    & 9(6)    & 12(5)   \\
                                   & $M_5$ \ \ W23            & 12(3)   & 12(3)   \\\hline
    \end{tabular}
    \caption{Inferred dipole amplitude by sample and model.
             All values are reported with a 95\% credible interval
             and given in units of $10^{-3}$.}
    \label{tab:amp_table}
\end{table}

\section{Discussion \& Conclusions}
\label{sec:discussion-conclusion}
\subsection{Effect of local clustering}
The results we presented in the previous section indicate that,
while the dipole in RACS-low and NVSS is consistent in direction
with the CMB dipole,
there is significant support for an amplitude larger
than that of the CMB dipole.
While the case is not decisive for NVSS(B) alone,
with the kinematic dipole being marginally preferred
over the other hypotheses,
RACS-low(B) has sizeable evidential support
for a dipole larger than the CMB expectation
($\mathcal{D}_{\text{CMB}} \approx 4.0 \times 10^{-3}$).
In addition,
the joint analysis -- either with the A or B variants -- favours
a larger dipole amplitude,
even if we neglect the results of \citet{wagenveld2023}
and consider model ($M_3$) to be the dominant explanation.

Nonetheless, one important observation is that the dipole amplitude
in the individual catalogues diminishes moving from the A to B variants.
Thus the clustering of low-redshift radio sources has a non-negligible impact on the dipole amplitude.
In fact, even though about 3000 sources were removed from NVSS and about 3700 sources from RACS-low, representing only $0.9\%$ and $0.8\%$ of the final source counts of each sample respectively, the inferred dipole amplitudes decreases by $\approx$10\%--15\%
depending on the sample used.
This accounts for why, in both B samples, the kinematic dipole
model gains a comparitively larger marginal likelihood
with respect to the other competing models (see Tables~\ref{tab:nvss_bayes} and \ref{tab:racs_bayes}).

\subsubsection{Clustering by redshift}
We can further probe the effect of local clustering on the dipole amplitude by noting how the amplitude changes with increasingly deeper redshift cuts.
The NED catalogue contains objects at a redshift of 0.01 and below, whereas 2MRS contains objects up to a redshift of about 0.1.
Although the homogeneity scale -- the distance over which the Universe is expected to average out to FLRW -- is presently thought to be smaller than that implied by $z \approx 0.1$ \citep[see e.g.][which place the transition to homogeneity around 100--115 Mpc ($z \approx 0.023$--0.027) assuming $h = 0.7$]{sarkar2009, scrimgeour2012}, there remains considerable uncertainty over its exact number \citep{aluri2023}.
It is therefore worth seeing how deeper cuts impact any clustering bias in the dipole amplitude.
To do this, we simply repeat our model fitting for different redshift bins.
In the first case, we remove no sources, and in the second case, we only remove sources with redshift $\leq 0.01$.
Then, we remove sources with $z \leq 0.02$, and so on.

\begin{figure}
    \centering
    \includegraphics[width=\columnwidth]{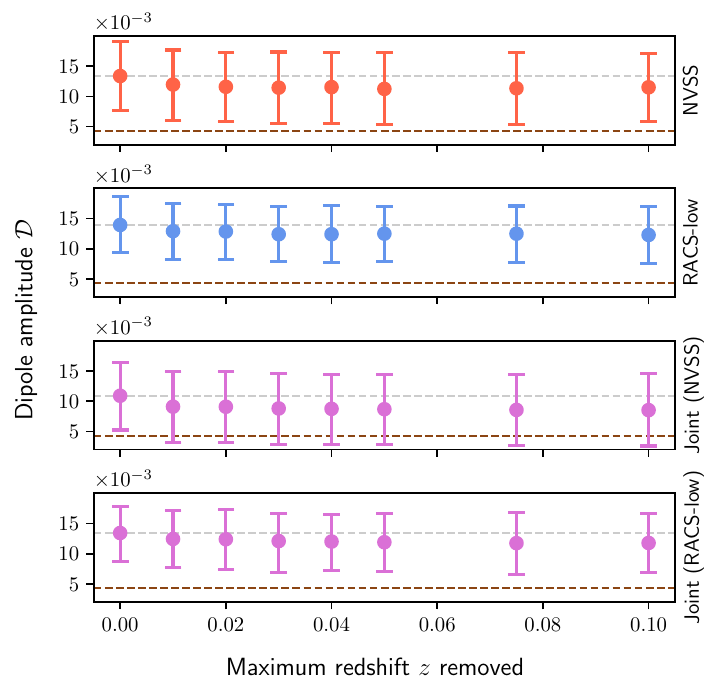}
    \caption{Inferred dipole amplitude against maximum redshift removed during cross-matching. The error bars represent a $95\%$ ($2\sigma$) credible interval. Red is NVSS, blue is RACS-low, and pink is the joint sample
    (the relevant samples are also labelled to the right of each pane). The brown line indicates the expected dipole amplitude in each case,
    and the grey line shows the inferred amplitude
    with no sources removed (i.e. the amplitude
    at $z = 0$).}
    \label{fig:amp_redshift}
\end{figure}

Our results for this process are presented in Fig.~\ref{fig:amp_redshift}.
Immediately, it can be inferred that the dipole amplitude continues to decrease up until $z \approx 0.04$ (comoving distance of $\approx170$ Mpc), where it starts to plateau off.
This is broadly the case whether or not
one considers the samples individually
or the joint analysis, in which both catalogues are assumed
to be affected by a dipole pointing in the same direction
but with different magnitudes.
We note that -- as mentioned earlier -- the uncertainties on the amplitude are quite significant.
None the less, by the $z < 0.03$ cut, the expected $\mathcal{D}_{\text{NVSS}}$ (shown by the dashed brown line
in the top pane of Fig.~\ref{fig:amp_redshift})
is just outside the $2\sigma$ uncertainties on the inferred dipole amplitude.
However, the disagreement between the expectation
and the inferred amplitude is much larger for RACS-low,
reaching $\approx 3.4 \sigma$
even after local sources have been identified and removed
from the sample.

\begin{figure}
    \centering
    \includegraphics{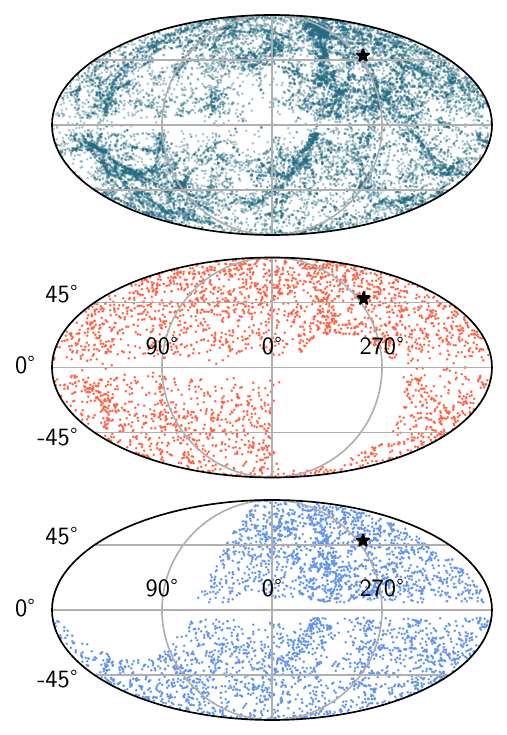}
    \caption{Spatial distribution (Galactic coordinates; Mollweide projection) of NED and 2MRS sources ($z < 0.02$) and those that are cross-matched in NVSS/RACS-low. The black star indicates the direction of the CMB dipole. Some sources may reside in masked NVSS/RACS-low regions since cross-matching is applied before masking. \textit{Top:} NED and 2MRS sources with redshift less than 0.02. The graticule labels have been removed for clarity. \textit{Middle:} All 3049 cross-matched NVSS sources removed in NVSS(B). \textit{Bottom:} All 3700 cross-matched RACS-low sources removed in RACS-low(B).}
    \label{fig:local_sources}
\end{figure}

To visualise the local structure being caught here, we plot the spatial distribution of NED and 2MRS sources which have a redshift lower than 0.02 in the top pane of Fig.~\ref{fig:local_sources}.
A significant band of low redshift structure concentrated near the CMB dipole can be seen.
This band of structure is being picked up in the cross-matched sources identified in NVSS and RACS-low, shown in the middle and bottom panes of Fig.~\ref{fig:local_sources} respectively.
This likely explains why the inferred dipole amplitude decreases until the redshift bin of $z < 0.04$; this local structure is being incrementally removed until the bin is sufficiently deep to cover its spatial volume.

With this established, it is worth mentioning the role that the NED sample has in our study.
Our chief aim is to remove as many low-redshift sources as possible.
However, since the NED database collates objects from independent studies and resources, there is the possibility that it introduces a bias or some other unknown effect in our results -- especially since those studies may have different levels of completeness and coverage.
As a consistency check, we repeated our above analysis without the NED sample and only the 2MRS sample, which, unlike the former, is constituted solely from the Two Micron All Sky Survey.
We find that our results are essentially unchanged with this modification, likely because the cross-matched NED sources represent only $\approx 10\%$ of the net cross-matched sources.
We can therefore say that our findings are not influenced by an unforeseen bias in the NED objects.
This supports our observation that clustering is a genuine influence on the dipole amplitude.

\subsubsection{Clustering dipole}
\begin{figure}
    \centering
    \includegraphics[width=\columnwidth]{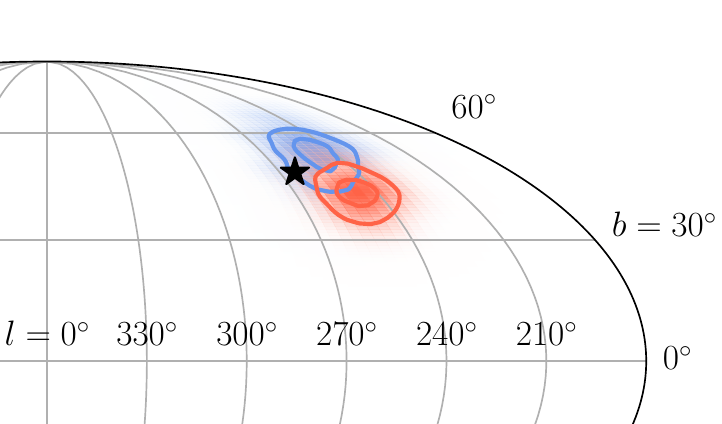}
    \caption{Projection of the marginal posterior distribution ($l$ and $b$) for the cross-matched source dipole onto the sky (Mollweide). Blue indicates the RACS-low clustering dipole and red the NVSS clustering dipole. The contours give intervals of $0.5\sigma$, encompassing 11.8\% and 39.4\%. Additional contours have been omitted for clarity. The black star indicates the direction of the CMB dipole.}
    \label{fig:cs_dipole}
\end{figure}

Another way of understanding the effect of local clustering is by visualising the dipolar contribution of local sources to the overall inferred dipole.
We refer to this as the `clustering dipole'.
More explicitly, we take the cross-matched sources for either NVSS or RACS-low -- as shown in the bottom two panes of Fig~\ref{fig:local_sources} -- and fit a free dipole (model $M_1$) to them.
For the cross-matched NVSS sources,
we infer a dipole amplitude of $\mathcal{D} \approx 0.25 \pm 0.06$,
and for the cross-matched RACS-low sources,
we infer an amplitude of $\mathcal{D} \approx 0.22 \pm 0.05$.
We project the resulting marginal posterior distribution for $l$ and $b$ onto the sky in Fig.~\ref{fig:cs_dipole} after performing this fit.
Strikingly, the cross-matched sources -- both for NVSS and RACS-low -- have a well-constrained dipole signal pointing near the CMB dipole.
A priori we expect that the clustering dipole might point in any direction on the sky, as it ultimately traces the net dipole component of nearby structure.
Meanwhile, the overall CMB dipole includes contributions from,
for example, the motion of our heliocentric frame of reference with
respect to the Galaxy.
This is not necessarily correlated with local structure.

However,
it is worth pointing out that
the CMB dipole -- i.e. the motion of the Sun with respect to the CMB -- points
in roughly the same direction as
the motion of the Local Group (LG) with respect to the CMB 
\citep[$l,b \approx 270^\circ, 30^\circ$; see Table 3 in][]{planck2020}.
Meanwhile,
the motion of our heliocentric frame with respect to the LG
is approximately oppositely-aligned with the CMB dipole.
Accordingly,
the fact that the clustering dipole seems to align with the CMB dipole
means that it is also roughly aligned with the axis defining the LG's motion
with respect to the CMB.
It is therefore possible that the kinematic and clustering
dipoles are correlated as implied in Fig.~\ref{fig:cs_dipole},
since the cause of the LG's motion with respect to the CMB
is gravitational interaction with local structure.
Even so, it is curious that the clustering dipole seems
to be well-defined with a small number of sources ($<4000$)
and aligns so closely with the CMB dipole.
At least for this study, it explains why the removal of local sources is correlated with a decreasing dipole amplitude; the clustering and matter dipoles act in concert, contributing to a larger inferred amplitude.
This is a 10\%--15\% effect, as was noted earlier.

The possibility of a local void or overdensity contributing to the NVSS net dipole amplitude
was considered in, for example, \citet{rubart2014}.
There, the authors simulated the effect of a perturbed region with source density
below the average ($\delta = - 1/3$) out to redshifts of 0.07,
arriving at a similar value of $\mathcal{D}_{\text{clust.}} \approx 2 \times 10^{-3}$
(although their approach is totally distinct from our own).
In a similar vein,
\citet{bengaly2019} constructed a forward-looking sample,
using predictions from $\Lambda$CDM to anticipate
the radio number count map
that will be made available by the upcoming Square Kilometre Array (SKA).
The authors found that the dipole estimate in this mock sample
is contaminated by local structure,
with most of the inconsistency between the true kinematic dipole
and the inferred dipole caused by sources out to $z = 0.1$.
Thus, they expect cross-matching radio SKA sources with optical/infrared
data to be of key importance in excising low redshift sources,
which is what we also conclude in this work.
In any case,
the two studies lend support to our proposition that local structure
contributes non-negligibly to the inferred dipole in NVSS and RACS-low.

\subsubsection{Joint analysis}
In our joint analysis,
we give results both where the catalogue A variants
and catalogue B variants are used (see Table~\ref{tab:joint_bayes}).
To reiterate,
the B variants have had the local structure in Fig.~\ref{fig:local_sources}
removed by crossmatching with both 2MRS and NED.
The results with the B variants are representative of our main finding.
In this case, 
we find that the model assuming the results of \citet{wagenveld2023}
as the prior likelihood function
yields the highest evidence by a significant margin.
If we blind ourselves to the findings of that work
and consider the next-best explanation of the data,
we find that fixing the dipole to the CMB direction
but allowing the amplitude to vary as a free parameter
offers the greatest predictive power.

\begin{figure}
    \centering
    \includegraphics[width=\columnwidth]{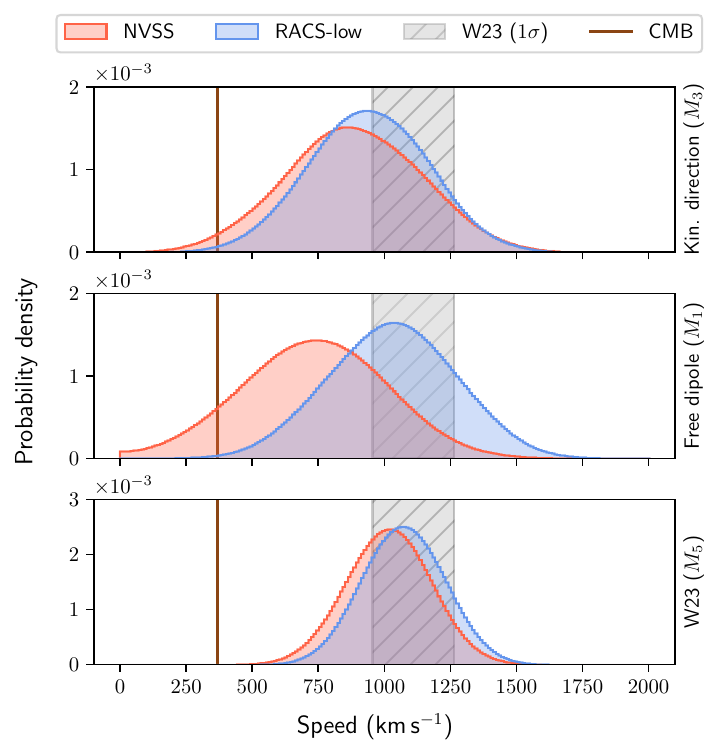}
    \caption{Marginal posterior distributions for the implied velocity
    of our heliocentric frame with respect to the CMB
    in the joint analysis (B variants).
    The distributions have been smoothed with a Gaussian kernel
    for the sake of visualisation.
    The top pane shows the inferred amplitudes assuming the
    kinematic direction model ($M_3$),
    while the middle and bottom panes assume the free dipole model ($M_1$)
    and the W23 model ($M_5$) respectively.}
    \label{fig:amp_comparison}
\end{figure}

We can understand this result by turning our attention
to the marginal distributions of the dipole amplitudes,
as shown in Fig.~\ref{fig:amp_comparison}.
There, one can see that by fixing $l$ and $b$ to the CMB dipole direction (top pane),
a dipole with larger amplitude than the CMB expectation (brown line) is preferred.
Models which fix $\mathcal{D}$ to the the CMB expectation are consequently penalised;
they give a poorer explanation of the data.

One interesting point is that, if we assume a free dipole ($M_1$),
the posterior distribution of the RACS-low amplitude tends to higher values
than for NVSS.
This is shown in the middle pane of Fig.~\ref{fig:amp_comparison},
although the effect is also discernible for the other models (see the other panes).
This is also present in our analysis of the individual samples, as shown in Fig.~\ref{fig:all_corners}.
It could therefore be asked why,
on net,
the RACS-low dipole tends to be larger than that NVSS --
both in the joint and individual analyses.
This cannot be explained by variation in the spectral index and/or flux distributions
of each sample,
since we convert to an implied velocity in Fig.~\ref{fig:amp_comparison}
using the dipole expectations.

It is possible that this amplitude discrepancy evinces further systematic effects biasing the amplitude which we have not been able to adequately account for in RACS-low.
For example, it is reasonable to suppose that we have not removed
all possible local radio sources in our cross-matching procedure.
This is especially the case considering the sensitivty of the amplitude to local clustering.
\citet{tiwari2016} estimate from mock NVSS catalogues that sources
with redshift $z < 0.1$ contribute to the dipole amplitude by as much as $70\%$. 
With a different methodology,
the fiducial model of \citet{cheng2023} estimates the clustering dipole
(the component of the inferred dipole arising from locally-clustered sources)
in NVSS to be $\approx (4 \pm 1) \times 10^{-3}$,
$80\%$ of which comes from objects below a redshift of 0.05.
This is about twice as large as the clustering component we inferred
in the foregoing sections.
Therefore, while the top pane of Figure~\ref{fig:local_sources}
might be a good representation of local structure,
it only provides a lower limit on the contribution to the clustering dipole;
there may be more sources which have not been accounted for.

\subsection{Comparison with other studies}
Keeping the effects of clustering in mind,
how does our result compare to other studies?
Speaking broadly, the current state of the literature leans towards a matter dipole pointing in the same direction of the CMB dipole -- which is consistent with our joint analysis --  but with a magnitude in excess of the CMB dipole amplitude \citep{Peebles_2022, aluri2023}.
Now, it is important to recognise that our amplitude estimates have significant uncertainties.
That being said, at face value
our results are consistent with prior observations of an excessive dipole amplitude.

For example,
our results are evidently consistent with \citet{wagenveld2023},
the prior knowledge of which offers the best explanation of the data.
Although,
we note that the
authors' approach differed slightly from ours,
since they assumed a common dipole amplitude across NVSS and RACS-low,
whereas we allowed it to vary between the samples.
In doing so,
we have made the additional observation that the RACS-low sample
prefers a larger dipole amplitude.
One other point of contrast is that the authors assumed
that the effects of clustering are negligible;
we have found substantial evidence that this is not the case.
Though accounting for clustering only slightly reduces the dipole amplitude
for the W23 model ($M_5$; $\mathcal{D} \approx 0.013 \to 0.012$),
it has a more pronounced effect on our other models.
We interpret this as evidence for clustering playing an important effect
in the \citet{ellis1984} test.

Turning to a separate study,
\citet{secrest2022} also examined NVSS,
finding an amplitude of $\mathcal{D} \approx (12 \pm 3) \times 10^{-3}$.
This result is encompassed by the uncertainties on our findings,
whether or not one turns to NVSS(A), NVSS(B) or the joint analysis.
The fact that the authors did not account for the clustering of local sources
can explain why their median amplitude is slightly larger than what we inferred
after cross-matching (see NVSS(B) in Table~\ref{tab:amp_table}).
We cannot comment on the joint case, since the authors compared the quasar catalogue of CatWISE2020 \citep{marocco2021} and NVSS.

Additionally, in \citet{colin2017},
the authors accounted for local clustering by cross-matching with 2MRS.
Nonetheless, this did not significantly impact their dipole amplitude estimate,
which was around four times greater than the expectation.
The lack of sensitivity of the dipole amplitude
to local sources conflicts with our findings.
Now, the authors cut and combined NVSS with SUMSS to create a merged catalogue
with $\approx 600\,000$ sources,
rather than considering both samples jointly
with a likelihood function like equation~\eqref{eq:joint_likelihood}.
Thus, it may be worth revisiting their work in the future
with a joint Bayesian approach and seeing if the contribution
from SUMSS still leaves an anomalously high dipole amplitude --
even when clustering is mitigated.

\subsection{Future outlook}
On the basis of the above, our result raises questions about the nature
of the tension between the CMB dipole and the matter dipole.
It sits in agreement with historic studies which have
reported an anomalously large radio galaxy dipole,
as well the result of \citet{secrest2021},
which confirmed this with quasars in the near-IR.
However, one outstanding issue
is why our finding is mildly in tension with that of \citet{mittal2024}.
One possible explanation is that,
with the additional dataset afforded by the joint analysis here,
the statistical power of the result increases -- noting 
that the analysis of \citet{mittal2024} dealt with Quaia alone.
Further, while the joint analysis points towards an excessive dipole amplitude,
there are still additional questions
regarding the full effect of local structure to pursue.
We therefore do not claim that the state of the tension
is unequivocal, clear-cut or well-understood.

Importantly, we have shown that the impact of clustering
in both NVSS and RACS-low is significant.
Contamination from local radio sources, 
even if they represent a small part of the sample,
disproportionately biases the inferred dipole amplitude to larger values.
This necessitates a more detailed profiling of local sources in future studies,
possibly including calculations of the clustering dipole expectation
as some authors have done previously.
We are therefore of the view that the common assumption
that clustering is negligible is no longer tenable,
and may at least in part account for a fraction of the dipole excess
reported in many radio galaxy studies.
One correlated question is why our inferred `clustering dipole'
(the dipole in the cross-matched sources)
appears to align with the CMB dipole.
Is this consistent with the predictions of the CP, or is it an anomaly?

With respect to our methodology,
there are important factors to consider about our cross-matching,
in which we pair near-infrared (2MRS) and radio (NVSS/RACS-low) sources.
Generally, cross-matching radio catalogues with other optical/infrared samples
is a non-trivial and well-known problem.
This is largely because radio sources are often extended
with multiple spatially distinct components,
and it can be unclear whether the radio emission is even associated
with the optical/infrared counterpart source itself
(\citealt{fan2020}; see e.g. \citealt{velzen2012} and \citealt{fan2020b}
for recent attempts at cross-matching manually and automatically respectively).
Roughly 10\% of radio sources will have extended,
spatially-separated lobes,
and it can be challenging to distinguish between these structures
(part of the same source) and genuinely distinct sources.
It is therefore conceivable that by cross-matching NVSS/RACS-low with 2MRS,
we occasionally only remove one of the many lobes
of a radio source and miss the extended structure.
In a follow up analysis,
we tried to probe this issue and remove highly-extended sources
by a cut in the ratio of the integrated to peak flux density
$S_{\text{int.}} / S_{\text{peak}}$.
We tried cuts from $S_{\text{int.}} / S_{\text{peak}} > 10$
to as low as $S_{\text{int.}} / S_{\text{peak}} > 5$,
but found that this has a negligible impact
on the inferred parameters and model marginal likelihoods.
It would therefore be worth exploring this in more depth
in future analyses.

One additional point relates to the actual value
of the cross-matching radius itself,
which we set at 5 arcseconds in light of the positional accuracy of NVSS/RACS-low objects.
A more thorough way of setting the radius would involve
constructing a `fake' catalogue with random positional offsets from NVSS/RACS-low,
then evaluating the fraction of genuine versus fake matches
as a function of radius \citep[see e.g.][]{mahoney2011, chhetri2020}.
This leaves scope for future work to investigate more robust
and consistent ways of removing local galaxies and handling the realities
of extended radio structure,
beyond what we have already achieved with our choice of mask and cross-matching.

Finally, our joint methodology invites extension with other catalogues.
The utility of this approach is that we may add
additional terms to equation~\eqref{eq:joint_likelihood} for each other sample,
and these need not be limited to radio galaxies.
For example, we could include samples of quasars or Type Ia supernovae.
Doing so improves the statistical power of the result by analysing independent,
unique objects in which different processing techniques have been used.
While an unaccounted for systematic effect could bias the result,
it would have comparatively less of an effect amongst the other samples in the joint analysis.
In future work, this will help further understand the nature of the claimed tension
between the CMB dipole and the matter dipole in the literature,
robustly interrogating the role of the CP in our cosmological paradigm.

\section*{Acknowledgements}
We thank the anonymous refereee for their insightful comments,
which greatly improved this paper.
We also thank Harry Desmond for helpful discussions
regarding the correlation between the kinematic and clustering dipoles.
Additionally, we extend our gratitude to Stefan Duchesne and Emil Lenc
for useful discussions regarding the RACS sample.
VM acknowledges Prof. Kulinder Pal Singh, IISER Mohali for his support; DST, Government of India for the INSPIRE-SHE scholarship and University of Sydney's internship program which made this work possible.
This scientific work uses data obtained from Inyarrimanha
Ilgari Bundara / the Murchison Radio-astronomy Observatory. We acknowledge the Wajarri Yamaji
People as the Traditional Owners and native title holders of the Observatory site. CSIRO’s
ASKAP radio telescope is part of the Australia Telescope National Facility
(\url{https://ror.org/05qajvd42}). Operation of ASKAP is funded by the Australian Government with
support from the National Collaborative Research Infrastructure Strategy. ASKAP uses the
resources of the Pawsey Supercomputing Research Centre. Establishment of ASKAP, Inyarrimanha
Ilgari Bundara, the CSIRO Murchison Radio-astronomy Observatory and the Pawsey Supercomputing
Research Centre are initiatives of the Australian Government, with support from the Government
of Western Australia and the Science and Industry Endowment Fund.
This research has made use of the NASA/IPAC Extragalactic Database (NED),
which is operated by the Jet Propulsion Laboratory, California Institute of Technology,
under contract with the National Aeronautics and Space Administration.

This work made use of the \textsc{python} packages
\textsc{dynesty} \citep{skilling2004, skilling2006, dynesty-v2.1.2}, \textsc{healpy} \citep{Gorski2005,Zonca2019},
\textsc{numpy} \citep{harris2020},
\textsc{matplotlib} \citep{hunter2007},
\textsc{pandas} \citep{mckinney2010,reback2020},
\textsc{scipy} \citep{scipy2020} and \textsc{astropy} \citep{astropy2022}.

\section*{Data Availability}
The ASKAP data analysed in this paper can be accessed through the CSIRO ASKAP Science Data Archive
(CASDA\footnote{\url{https://data.csiro.au/domain/casdaObservation}}) under project
code AS110.  
Other data used in this study will be made available with a reasonable request to the authors.



\bibliographystyle{mnras}
\bibliography{biblio} 








\bsp	
\label{lastpage}
\end{document}